\documentclass[fleqn,usenatbib,15pt]{mnras}
\usepackage{amssymb,amsmath}
\usepackage[T1]{fontenc}
\usepackage{ae,aecompl}
\usepackage{subfigure}
\usepackage{graphicx}	% Including figure files
\usepackage{amsmath}	% Advanced maths commands
\usepackage{amssymb}	% Extra maths symbols
\usepackage{color,xcolor}
\usepackage{newtxtext,newtxmath}
\usepackage[T1]{fontenc}
\DeclareRobustCommand{\VAN}[3]{#2}
\let\VANthebibliography\thebibliography
\def\thebibliography{\DeclareRobustCommand{\VAN}[3]{##3}\VANthebibliography}

\def\msun{M_\odot}
\def\kms{km s$^{-1}$}
\def\klms{{\rm km s^{-1}}}
\def\cc{{\rm cm^{-3}}}
\def\sc{{\rm cm^{-2}}}

\def\micron{{\rm \mu m}}
\def\htwo{{\rm H_2}}
\def\nht{{\rm NH_3}}
\def\nthp{{\rm N_2H^+}}

\title[Filament Intersections and Cores]{Filament Intersections and Cold Dense Cores in Orion A North}

\author[Chao Zhang et al.]{
Chao Zhang$^{1}$,
Zhiyuan Ren\thanks{E-mail: renzy@nao.cas.cn}$^{1}$,
Jingwen Wu\thanks{E-mail: jingwen@nao.cas.cn}$^{1}$,
Di Li$^{1,2,3}$,
Lei Zhu$^{7}$,
\newauthor
Qizhou Zhang$^{5}$,
Diego Mardones$^{6}$,
Chen Wang$^{4}$,
Hui Shi$^{1}$
Nannan Yue$^{1,2}$,
\newauthor
Gan Luo$^{1,2}$,
Jinjin Xie$^{1,2}$,
Sihan Jiao$^{1,2}$,
Shu Liu$^{1,2}$,
Xuefang Xu$^{1,2}$,
Shen Wang$^{1,2}$
\\
$^{1}$ National Astronomical Observatories, Chinese Academy of Sciences, A20 Datun Road, Chaoyang District, Beijing 100101, China \\
$^{2}$ University of Chinese Academy of Sciences, Beijing 100049, China \\
$^{3}$ NAOC-UKZN Computational Astrophysics Centre (NUCAC), University of KwaZulu-parental, Durban 4000, South Africa \\
$^{4}$ CSIRO Data61, Sydney, NSW 2015, Australia \\
$^{5}$ Center for Astrophysics | Harvard \& Smithsonian, 60 Garden Street, Cambridge, MA 02318, USA \\ 
$^{6}$ Departamento de Astronomia, Universidad de Chile, Casilla 36-D, Santiago, Chile \\
$^{7}$ Chinese Academy of Sciences South America Center for Astronomy
}

\date{Accepted XXX. Received YYY; in original form ZZZ}

\pubyear{2019}

\begin{document}
\label{firstpage}
\pagerange{\pageref{firstpage}--\pageref{lastpage}}
\maketitle

\begin{abstract}
We studied the filament structures and cold dense cores in OMC-2,3 region in Orion A North molecular cloud using the high-resolution $\nthp$ (1-0) spectral cube observed with the Atacama Large Millimeter/Submillimeter Array (ALMA). The filament network over a total length of 2 pc is found to contain 170 intersections and 128 candidate dense cores. The dense cores are all displaced from the infrared point sources (possible young stars), and the major fraction of cores (103) are located around the intersections. Towards the intersections, there is also an increasing trend for the total column density $N_{\rm tot}$ as well as the the power-law index of the column-density Probability Distribution Function (N-PDF), suggesting that the intersections would in general have more significant gas assembly than the other part of the filament paths. The virial analysis shows that the dense cores mostly have virial mass ratio of $\alpha_{\rm vir}=M_{\rm vir}/M_{\rm gas}<1.0$, suggesting them to be bounded by the self gravity. In the mean time, only about 23 percent of the cores have critical mass ratio of $\alpha_{\rm crit}=M_{\rm crit}/M_{\rm gas}<1.0$, suggesting them to be unstable against core collapse. Combining these results, it shows that the major fraction of the cold starless and possible prestellar cores in OMC-2,3 are being assembled around the intersections, and currently in a gravitationally bound state. But more extensive core collapse and star formation may still require continuous core-mass growth or other perturbations. 
\end{abstract}

% r0, NH2, sigma,  


% star-formation efficiency: may be overestimated due to missing flux (Hacar2018): 
% Kelvin / Jy = ???????
% decline of thermal and turbulent energy at the intersections.
% ++ kirk2017 cores: r>0.017 pc = 8'', the cores are apparently  embedded
% virial analysis decision:
% 1. 交叉点周围密度均匀，所以用球对称形式
% 2. 只要Pic足够大，有些core即使subvirial,也能塌缩。
% Pic following Konyves2015 - McKee Tan 2013
% Konyves2015: Sigma_BE/Sigma_BG > 1.5 may be checked.
% in deriving the filament PDF, the intersection areas are subtracted.
% shimajiri19 is an obs paper.

\section{Introduction}
\label{sec:intro}
Filaments are widely existing structures in molecular clouds and are closely related to star-forming activities \citep{schneider79, wang08, andre10, arzoumanian11}. Filamentary clouds usually have hierarchical structures \citep{hacar13, takahashi13, clarke16, gomez18} scaled from 0.01 to several parsecs. The theoretical works expect filamentary structures to be responsible for stabilizing the dense-gas and channeling gas into the dense cores or young stellar clusters \citep{pon11,tan14,smith16,motte18,lin19}, and in some cases more intensely support the mass aggregation via the the intersected and merged multiple filaments \citep[e.g.][]{hill11, myers11, hennemann12}. 

The observational studies have also revealed the universal connection between the filamentary networks and the dense cores and young stars \citep{arzoumanian11, schneider12, ragan12, henshaw16, kainulainen17, lin17, xu18}. In some regions, the major filaments in a cloud are connected into a filament-hub system and exhibit possible convergent gas motion towards the hub-area \citep{hill11, hennemann12, chen19b, tmorales19}. In other cases, the filaments are resolved into sub-structures at smaller scales, and the intersected subfilaments are closely associated with multiple or clustered cores or young stellar objects (YSOs) \citep[e.g.][]{schneider12, hennemann12, henshaw16, lin17, shimajiri19}, suggesting increased star-forming activities therein. 

% The magnetic field is also revealed to significantly influence the gas distribution in filament structures \citep[e.g.][]{liu18,wang19}. The fragmentation of the parental filament leading to core formation is also observed in a number of regions \citep{andre10, molinari14, fontani18}. 

Despite the spatial association, the physical connection between filaments and dense cores, in particular those at early stages, are still to be further explored. In doing this, one may first need to enlarge the sample of filament structures wherein the individual prestellar cores can be resolved to make direct comparison with their parental filaments. Second, in addition to the spatial comparison, it further requires a specified study about the key properties of the filaments including their spatial structure, velocity field, and magnetic field, in order to compare their relative importance in the dense core formation. Currently such detailed studies have been performed only in a limited number of regions \citep[e.g.][]{liu18,wang19}. 

% The stability of the cores by comparing the binding (self-gravity, external pressure) and supporting (thermal and turbulent motions and magnetic field) energies.

Orion A cloud is one of the most favorite sites to study the cold dense filaments and cores \citep{johnstone99,nutter07,sadavoy10,shimajiri15,kirk17,hacar18}. Its main body is an Integral-Shaped dense Filamentary cloud (denoted as ISF) elongated over the Orion Nebular Cluster (ONC) from North to South. The cloud should have a distance comparable to the ONC that is $D=388\pm 5$ pc \citep{kounkel17}. In recent work, using the ALMA+IRAM 30m combined data of $\nthp~J=(1-0)$ transition and the HiFIVE algorithm, \citet[][H18 hereafter]{hacar18} analyzed the dense gas in Orion A, revealed the peculiar fine structure of intertwined fibers within major filaments, and discussed their evolutionary state regulated by the gas density.  A series of relevant works were also addressed to study the dense clumps and young stellar objects (YSOs) therein \citep{shimajiri15,salji15,dario16,kirk17}. With general physical conditions and YSO catalogue well documented, it is desirable to further study the internal filament structures around the individual dense cores, thereby contributing an updated sample to demonstrate their mutual evolution.    
 
In this work, we present an observational study of the filaments in Orion A North (OMC-2,3) using the high-resolution $\nthp$ $J$=(1-0) transition observed with the Atacama Large Millimeter/submillimeter Array (ALMA).  The $\nthp$ (1-0) line has a high critical density of $1.5\times 10^5$ cm$^{-2}$ and a low degree of depletion in cold dense gas due to its low tendency of adhering to particles \citep{bergin97, pagani09, caselli02, miettinen14}. It is thus continuously adopted (following H18) in the current work to study the gas structures around the individual cores within the ISF. With high angular resolution reaching 1000-AU scale, the parental cores of individual YSOs can be resolved \citep[e.g.][]{ren14, kainulainen17, matsushita19}. Orion A cloud contains a large fraction of dense gas still in a cold and quiescent state \citep{salji15, kainulainen17, hacar18}, thus provides an ideal sample to study the initial evolutionary state of the dense cores. In addition, extensive infrared surveys are carried out to identify the YSOs and clusters in this cloud \citep{megeath12,salji15,shimajiri15}, so that the evolutionary states of the molecular cores, in particular the pre- and protostellar ones, can be classified from their spatial association with the catalogued YSOs. 

In this work we used several different methods to quantify the gas assembly in the filaments. First, the dynamical models suggested that the column-density probability distribution function (N-PDF) can estimate the turbulence decay and density increase due to the self-gravity \citep{krumholz05,hennebelle11,padoan11}. The N-PDF properties, including its column density range, power-law tail, and the spatial variation, can give a quantitative estimate for the dense-gas assembly and star-forming efficiency \citep[e.g.][]{federrath13,chen18}. Second, we calculated the virial and critical masses of the cores \cite{li13}, which can measure their gravitational binding state and tendency to collapse, respectively. 

% binding (self-gravity, external pressure) and supporting (thermal and turbulent motions and magnetic field)  

The following contents of this paper are organized as follows: In Sec.\,\ref{sec:data}, we introduce the observations and data reduction, and describe the algorithm used to extract the filament structures from the line emission data cube. In Sec.\,\ref{sec:result}, we describe the measurement of the gas distribution along the filaments and around the intersections, and the estimation of the core stabilities. The evolutionary stage and the trend of star formation of these gas components are discussed in Sec.\,\ref{sec:discussion}. The major findings are summarized in Sec.~\ref{sec:summary}.

\section{Data Reduction and Processing}\label{sec:data}
\subsection{Data and Observation}
The ALMA observation of the $\nthp$ (1-0) transition in OMC-2,3 (proposal ID 2013.1.00662.S, PI: Diego Mardones) was carried out during November 2014 and August 2015 in Band 3. Four bands were used, one of which covered the $\nthp$ (1-0) transition at $f=93.173$ GHz with a channel width of $\Delta V_{\rm chan}=0.1$ \kms. We used the 12-m main array and the 7-m Atacama Compact Array (ACA) to perform a mosaic mapping.  There are 14 pointings towards the OMC-2 and 11 pointings towards OMC-3. The pointing centers and on-source integration time are presented in Table 1. In observation, the 12-m array is at the nearly most compact configuration, with the shortest baseline of 5.2 $k \lambda$, corresponding to a spatial coverage up to 29 arcsec. The 7-m array (ACA) has a fixed configuration with spatial coverage of 13-73 arcsec. The entire data cube thus has a continuous coverage from 3-73 arcsec, which is sufficient to detect the filament structures in individual gas clumps, which have diameters of $2 r \leq 0.1$ pc \citep{kirk17}.  

% (175 minutes on-source time) on the OMC-2 region and the other seven EBs (174 minutes on-source time) on the OMC-3 region. There are 13 EBs (340 minutes on-source time) for OMC-2 region and three EBs (97 minutes on-source time) for OMC 2 region by using ACA observation. 

For the 12-m array data, we used J0423-0120, J0501-0159 and J0750+1231 as bandpass calibrators, J0517-0520 as phase calibrators, and J0423-013 as the flux calibrator. For the ACA data, we adopted J0501-0159 and J0607-0834 as the bandpass calibrator, J0541-0541 and J0607-0834 as phase calibrators, and Callisto and J0423-0120 as flux calibrators. The observational data were manually calibrated for each EB with CASA 4.4.0. We combined the UV datasets of the 12-m array and ACA and CLEANed it in CASA using the interactive model. The restored image was corrected for the primary-beam response. The synthesized full-width half-maximum (FWHM) beam size is 3 arcsec $\times$ 3 arcsec. The rms level of the final maps is 7 mJy/beam at a spectral resolution of 0.1 km/s, corresponding to $T_{\rm b}=0.11$ K.

\subsection{Extracting the filamentary structures in PPV space} \label{subtables} % 2.2
In previous studies, various algorithms have been applied to analyze the gas structures in molecular clouds, including Curvelet Transform \citep{starck03}, Filfinder \citep{Filfinder}, and HiFIVE \citep{hacar13}. Curvelet Transform can enhance the input elementary structure, especially the elongated features in the data, but it cannot extract filaments in spectral data cube in the Position-Position-Velocity (PPV) space. Filfinder is capable of uniformly extracting the hierarchical filament structures. It can have stable performance even if the image has large intensity variation, but like the curvelet Transform, it also deals with two dimensional (2D) images. HiFIVE is developed to process the 3D data cube. As shown in H18, it is powerful in resolving the gas structures in PPV space. It can identify each velocity-coherent gas structure in the PPV space, and will clearly divide the different structures. These structures can be further modeled using one dimensional paths, such as filaments and fibers. 

Another algorithm for processing 3D data cube is the Discrete Persistent Structures Extractor (DisPerSE, \citealt{sousbie11}). This algorithm can sensitively identify the elongated gas structures in any 3D data cube above the adopted intensity threshold. It has been commonly used in cosmic dataset to identify the filaments and voids. DisPerSE requires a filament to go through one or more saddle points. At a saddle point, there are two particular directions along which the intensity profiles rightly have local maximum and local minimum at this point, respectively. For a saddle point in the filament, the two directions are usually along and perpendicular to the filament, respectively. DisPerSE requires each filament path to contain at least one saddle point. An isolated and elongated single dense core has no internal saddle point thus would not be misidentified as a filament. The extracted filament structures are stable against the the intensity variation bellow the intensity threshold.  

Both HiFIVE and DisPerSE can model the gas structures in PPV data cube. Since HiFIVE has been already applied to Orion A in H18, we adopted DisPerSE in this work. DisPerSE would first extract the individual filament paths above the threshold and allow us to inspect their intersections at the next step. Following the previous studies \citep[e.g.][]{sousbie11,arzoumanian11,schneider12}, we adopted the $5\sigma$ level as the threshold. In our data cube, it corresponds to $5 \sigma=35$ mJy beam$^{-1}$ in the current data set. 

After the filament extraction, we examined the possible misidentifications. The misidentified filament paths would go through (i) separated cores or (ii) two dimensional sheet-like structures. We found that the entire filament structure is located in the emission region above the $5\sigma$ limit, while the isolated cores are not covered by the filaments. Meanwhile, the filament paths are measured to have full widths in a narrow range of $w_{\rm fil,full}=14\pm 4$ arcsec above $5\sigma$ (or FWHM widths of $w_{\rm fil,FWHM}$=$7\pm2$ arcsec) and line widths of $\Delta V=0.5\pm 0.2$ \kms, suggesting that the paths should trace the dense gas elongation in one dimension, while the sheet-like morphologies are rare.     

% Because our observed region is not wide enough to cover the filaments located away from the ISF, we only extracted the filaments on the ISF. The filaments extending beyond the observed region are not included. 

% A few weak gas patches isolated from the ISF have different $w_{\rm fil}$ and $\Delta V$. are thus not considered in our study. The identified intersections are also denoted with yellow dots in filament structures (Fig.\,\ref{fig:3dfilament}).

\section{Result} \label{sec:result}
\subsection{Filaments and intersections: overall properties} 
The integrated $\nthp$ (1-0) emission and the filament paths extracted by DisPerSE are displayed in 5 sub-areas. Each sub-area contains a frame of relatively discrete gas structure. Fig.\,\ref{fig:omc23_ppv}a shows the entire emission region in OMC-2,3 overlaid on the Spitzer/IRAC 8 $\micron$ continuum emission. Fig.\,\ref{fig:omc23_ppv}b shows the 3D structure of the filament paths in Area-1. In Fig.\,\ref{fig:omc23_ppv}c, the filament paths are plotted together with the $\nthp$ emission in the PPV data cube, wherein the hyperfine component of $\nthp$ $(1_0-0_1)$ is adopted to plot and analyze the gas structures in the PPV space.  This component is isolated from the other HFCs (see Fig.\,\ref{fig:spectra_fitting}) and would most clearly demonstrate the velocity distribution of the cold dense gas. 

The $\nthp$ (1-0) emission in the sub-areas are shown in Fig.\,\ref{fig:area_maps}. The filament paths projected on the plane of the sky are displayed in yellow lines. The intersections and local emission peaks are labeled with circles and plus symbols, respectively. Based on the observed gas distributions and intensity variation over the filaments,  we posed a few criteria in selecting and denominating the typical gas structures:

(1) an \textit{intersection} represents a position where three or more filament paths are connected;

(2) a \textit{filament path} represents a segment between two adjacent endpoints. An endpoint could be either an intersection or an isolated terminal; 

(3) a \textit{branch} represents a single filament path between an intersection and a terminal;

(4) a \textit{local emission peak} should exceed the surrounding emission intensity above 5$\sigma$ intensity (35 mJy beam$^{-1}$ or $0.35\times 10^{23}\,\sc$). The surrounding level is adopted as the average emission intensity at $r$=7 arcsec, satisfying $2 r$=$\overline{w}_{\rm fil,full}$=$14''$. This is based on the consideration that the dense gas assembly usually has comparable or smaller spatial extents than the filament width.

(5) a \textit{candidate core} is selected from the local emission peaks if it is dominated by a single velocity component in its $\nthp$ spectrum. The single velocity component guarantee that the emission peak is more likely tracing a real mass assembly instead of overlapped filaments. The exclusion of overlapped filaments are described in detail in Sec.\,\ref{sec:sec_pv}. 

% The JCMT/SCUBA-II 850 $\micron$ continuum emission is also overlaid in dashed contours. Several representative long single-path filaments are labeled with blue-dashed arrows.

As shown in Fig.\,\ref{fig:area_maps}, in general, the output filament paths are closely associated with the observed emission features. Their relation can be described in three major aspects: (i) the identified filament paths go through all the emission features above the $5\sigma$ detection limit; (ii) the spatial distribution of the filament paths are coherent with the spatial extents of the emission features; (iii) the major fraction of the emission regions have intersected filaments with relatively short filament paths ($l<20$ arcsec). In contrast, a few very elongated structures ($l\geq20$ arcsec) are identified as long filament paths, as labeled in Fig.\,\ref{fig:area_maps}. 

As shown in Fig.\,\ref{fig:area_maps}, the filament intersections are widely distributed over the observed region. Each intersection often have three or four filament paths, including one branch and other two or more paths that are connected to other intersections. In some cases, the paths from one intersection are all connected to other intersections. As a result, these filament paths could become encircled, forming a ring-like morphology. A typical ring in Area-3 is labeled with the dashed box in Fig.\,\ref{fig:area_maps}. A series of more complicated multiple rings can be seen in Area-1, in particular MMS-2,3 region. This region was revealed to have complicated hierarchical structure of dense cores aligned over the filament \citep{takahashi13}. The fragmentation and dense core formation could be influenced by the closely intersected filament rings. 

The filament structures are compared with fibers identified by H18, as shown in Fig.\,\ref{fig:fila_compare}. The left panel shows the fibers in H18, which extend over OMC-1,2 and a part of OMC-3. The right panel shows the fibers overlaid in OMC-2 FIR-2,3,4 region, which is covered in both H18 and this work. As the figure shows, for each fiber that has overlap with the $\nthp$ filaments, its direction is in parallel with the bulk gas elongation of the  filament paths. But the individual filament paths are mostly shorter than the fibers, and the exact overlap between the fibers and filament paths is rare. This is within our expectation since the fibers would tend to trace large-scale structures because of two reasons. First, the $\nthp$ image in H18 also include a component from the single-dish data (IRAM 30-m), which mostly trace the spatial scales above 30 arcsec (IRAM 30-m beam size), Second, the fiber morphology also depends on HiFIVE algorithm. HiFIVE would first cover the entire area of each velocity-coherent structure, then extract the principal axis of the structure as the fiber direction. In this process, the fiber would mainly delineate the overall extension of the structure, and be less sensitive to the internal variations (scale $<30$ arcsec). 

% These branches may act as a channel of transporting gas from diffuse gas onto the major filaments \citep[e.g.][Figure 9 therein]{andre14}.  

\subsection{Intersections: Velocity Distribution} \label{sec:sec_pv} 
We examined the gas distribution and kinematical features around the intersections based on their velocity distributions. First, some partly overlapped filaments could be actually separated along the line of sight, the overlapped areas are thus not real intersections. We can eliminate such ``pseudo-intersections" based on the velocity distribution. One can assume each filament to have a distinct velocity, so that if two filaments are separated in the PPV space, they are also likely separated along the line of sight. This assumption is adopted by other works to identify the intertwined filaments or fibers \citep[e.g.][]{hacar13,shimajiri19}. 

The second scenario to be cautioned is the mass transfer flow along the filament. The gas flow can also generate large velocity variation near the intersections \citep[e.g.][]{peretto14}. But the transfer flow is characterized by a sudden velocity change and around the center, which indicates the gas flow being accelerated and halted at the central YSO. In contrast, the overlapped velocity components would have more separated velocities and less steep velocity gradient, so that each velocity component would extend smoothly along the PV-cut direction, instead of having a drastic variation over the center.  

The three typical cases, including an intersection with single velocity component, a pseudo intersection, and an intersection with large velocity gradient, are shown in three columns in Fig.\,\ref{fig:section_example}. In each column, the upper panel shows the filament paths overlaid on the integrated $\nthp$ emission, the middle and lower panels show the position-velocity (PV) diagrams along the possible gas elongation (major axis) and the transverse direction (minor axis) as denoted in the upper panel. 

In the example of one single velocity component (left column), the three filaments have a consistent systemic velocity of $V_{\rm sys}=11.4$ \kms\ at the intersection (upper panel). The PV diagrams (middle and bottom panels) show that the bulk of line emission is confined within a velocity range of $V-V_{\rm sys}=\pm 0.3$ \kms, which is comparable to the average line width of $\overline{\Delta V}=0.5$ \kms\ of the $\nthp$ emission region, as labeled with red dashed lines. 

Around the pseudo intersection (middle column), the overlapped filaments with different velocities can be discerned from their paths (upper panel), and the components at different velocities ($V-V_{\rm sys}=0$ and 0.6 \kms) have comparable intensities at the central position (offset=0) as labeled with dashed lines in the PV diagrams. 

For the intersection with large velocity gradient (right column), the PV diagrams also exhibit much broader velocity distribution than $\overline{\Delta V}$, but the multiple velocity components are not overlapped at the center. Instead, they becomes converged into a relatively narrow range that is also comparable to $\overline{\Delta V}$. 

The comparison among the three cases suggests that one can use $\overline{\Delta V}$ as a threshold to distinguish the simple velocity component from more complicated cases. If an intersection has velocity variation significantly exceeding $\overline{\Delta V}$ in its surroundings, it would be suspected to have multiple components.  According to this criterion, we selected 156 intersections with simple velocity distributions ($\Delta V \leq \overline{\Delta V}$) and 34 intersections that have large velocity variations. Among these intersections, 14 ones are found to be possible pseudo intersections. The other 20 ones may have largely velocity gradients instead of multiple components. So altogether we have 176 intersections that are considered to be real. 

We note that the selection based on $\Delta V$ would still be insufficient to exclude all the overlapped filaments. The exceptional ones would represent overlapped filaments with nearly equal $V_{\rm sys}$ so their line profiles are not evidently distinguished. We suppose this scenario to be scarce because the different filaments usually have noticeably different velocities ($>0.3$ \kms). If an intersection has a single-peaked spectrum and have relatively narrow line width ($\Delta V < \overline{\Delta V}$), its parental filaments are more likely to be indeed merged in space, while a pseudo-intersection would tend to have multiple spectral components.        

% cores are also inspected for multiple velocity components. 
% H18 also studied the cold dense gas in Orion A based on the $\nthp$ (1-0) image, and analyzed the filamentary gas structures in Orion A. The elongated gas structures are modeled by a series of fibers. The H18 fibers are      
% The fibers thus mainly delineate the structures larger than 20 arcsec. In this sense, the currently identified filament paths should represent the internal structures of the fibers, with spatial scales from 3 to 25 arcsec.

% cores: area1=22+4,  area2=25+4,  area3=22+10, area4=24+4, area5=13+3
% 22+25+22+24+13=103  4+4+10+4+3=25

\subsection{Dense Cores: Column Density, Mass, and Spatial Distribution}
As described in Section~\ref{sec:intro}, a key property to explore is the gas assembly and dense core formation in the filaments. We inspected the $\nthp$ emission at the local emission peaks (candidate dense cores). We also examined the spectra at the emission peaks to exclude those with evident multiple velocity components. The excluded ones are mostly around the pseudo-intersections.  

Over the ISF, we altogether identified 128 candidate cores that have single spectral component and exceed the $5\sigma$ detection limit. Their locations are labeled with plus symbols in Fig.\,\ref{fig:area_maps}. We compared their spatial distributions with the filament intersections. As a result, 119 intersections (70 \%) are found to have nearby cores within a distance of 5 arcsec. Inversely, among the dense cores, 103 (79 \%) are located close to the intersections. The other 25 cores are located on the long filament paths or displaced from the filaments. These cores are labeled with red crosses in Fig.\,\ref{fig:area_maps}. It is noteworthy that the intersections and the nearby cores usually have an offset of 3 to 5 arcsec and are only occasionally fully overlapped. This indicates that the filament extraction by DisPerSE is not biased to the dense cores.         

To examine the physical properties of the cores, we first derived the total column density $N_{\rm tot}=N({\rm H_2+HI})$ from the $\nthp$ emission following the normal procedure \citep[e.g.][]{caselli02, henshaw14}:

\begin{equation}\label{equ:nthp_ncol} % equation 1 in henshaw2014
\begin{aligned}
N_{\rm tot}  = & \frac{I_{\rm tot}}{X(\nthp)} \frac{8 \pi}{\lambda^3 A_{\rm ul}} \frac{g_l}{g_u} \times \frac{1}{J(T_{\rm ex})-J(T_{\rm bg})} \\
              & \times \frac{1}{1-\exp(-h\nu /k_{\rm B} T_{\rm ex})} \times\frac{Q_{\rm rot}(T_{\rm ex})}{g_l \exp(-El/k_{\rm B} T_{\rm ex})},
\end{aligned}
\end{equation}
wherein $I_{\rm tot}$ is integrated intensity of the $\nthp$ (1-0) line, $X(\nthp)$ is the $\nthp$ abundance, $A_{\rm ul}=3.63\times 10^{-5}$ s$^{-1}$ is Einstein coefficient \citep{schoier05}, $g_u$ and $g_l$ are the statistical weights of the upper and lower states, respectively; $J_\nu(T_{\rm ex})$ and $J(T_{\rm bg})$ are equivalent Rayleigh-Jeans excitation and background temperatures, respectively; $Q_{\rm rot}(T_{\rm ex})$ is the partition function at the excitation temperature $T_{\rm ex}$. The intensity $I_{\rm tot}$ can be derived from the optically thin component $JF_1=(1_0-0_1)$ using $I_{\rm tot} = 1/R_i \int T_{\rm b} {\rm d}v$, where $R_i=0.11$ is its relative statistical weight. 

For the $\nthp$ abundance, we also referred to the latest measurement in H18 that is $X(\nthp) = (7.5\pm 7)\times 10^{-10}$. The spatial variation of $X(\nthp)$ should be mainly caused by stellar heating. This effect can be evidently seen in two typical regions, OMC-3 MMS-5 and OMC-2 FIR-4 (Fig.\,\ref{fig:area_maps}), wherein $\nthp$ emission becomes largely devoid around the YSOs. Except these areas, the $\nthp$ emission is very coherent with the SCUBA 850 $\micron$ emission, suggesting that $X(\nthp)$ should have a uniform distribution. Since the current work is mainly focused on the cold dense gas component, it should be reasonable to assume a relatively stable level of $X(\nthp)$.      

The excitation temperature $T_{\rm ex}$ can be estimated from fitting the hyper-fine structure of the $\nthp$ spectra. We performed spectral fitting using the Python package Pyspeckit\footnote{Toolkit for fitting and manipulating spectroscopic data in python.}, which can generate a theoretical spectrum using the radiative transfer functions. The free parameters include $T_{\rm ex}$, optical depth $\tau$, systemic velocity $V_{\rm sys}$, and the velocity dispersion $\sigma_v=\Delta V/\sqrt{8\ln 2}$. An example of spectral fitting is shown in Fig.\,\ref{fig:spectra_fitting}a. One can see that the observed spectrum can be closely reproduced by adjusting the model parameters. Over the $\nthp$ emission region, the temperature is measured to be $T_{\rm ex}=15\pm 4$ K (Fig.\,\ref{fig:spectra_fitting}b). It can be compared with the kinetic temperature $T_{\rm kin}$ measured from the $\nht$ lines \citep{li13,kirk15}. These two works provided similar values of $T_{\rm kin}=16$ to 27 K in Orion A. Their lower limit is close to the current value of $\overline{T}_{\rm ex}(\nthp)=15$ K. Since the $\nthp$ should usually trace cooler and denser gas components than $\nht$ \citep{shirley15,chen19a}, it should be reasonable to adopt a constant $T_{\rm ex}=15$ K in calculation. It is also comparable to the average gas temperature of 15 K in infrared dark clouds \citep{chira13}. The $\nthp$ ranges of the filaments and cores are presented in Table 2.   

We measured the spatial extent of each core from its $N_{\rm tot}$ profile over the emission peak. The $N_{\rm tot}$ distributions along the major and minor axes of each core are overlaid in Fig.\,\ref{fig:nh2_profile} (upper panels). Along the minor axis, the $N_{\rm tot}$ profiles rapidly decrease towards the outer part (offset=$\pm7''$) and reaches the surrounding level of $\overline{N}_{\rm tot}=0.7\times 10^{23}\,\sc$. Along the major axis, the $N_{\rm tot}$ profiles are much more flattened, decreasing to the level of $\overline{N}_{\rm tot}=1.5\times 10^{23}\,\sc$ at offset=$\pm 12''$. The spatial extent of a core can be characterized by its major and minor axes. The cores are measured to have diameters of $2 r_{\rm maj}=8.5\pm3.0$ arcsec ($3300\pm 1100$ AU) and $2 r_{\rm min}=7.5\pm3.0$ arcsec ($2910\pm 1100$ AU), which are comparable to the filament width $\overline{w}_{\rm fil,FWHM}$=$7''$. The major and minor axes are nearly parallel with and perpendicular to the local filament path, respectively (see Fig.\,\ref{fig:section_example}). The morphologies of the cores agree with the expectation that they are being formed within the filaments.

We calculated the core mass from its integrated flux using $M_{\rm core}=\mu m_{\rm H} \int N_{\rm tot} {\rm d}A $, wherein $m_{\rm H}$ is the molecular hydrogen mass and $\mu=2.33$ is the mean molecular weight \citep{myers83}. The mass of the filament paths are also estimated from the flux within the average full width of the filament paths ($\overline{w}_{\rm fil,full}=14''$). The derived mass scales are presented in Table 2.  We note that in calculating the core mass, the surrounding levels of $N_{\rm tot}$ (solid horizontal lines in Fig.\,\ref{fig:nh2_profile}a, b) are not subtracted. This is based on the consideration that the mass assembly in the cores is taking place on the basis of the surrounding gas density, thus should exclude the contribution of the surrounding level. Although the surrounding level could also include the diffuse gas in the back- and foreground, this component would be insignificant because the average $N_{\rm tot}$ away from ISF is very low (see Fig.\,\ref{fig:npdf_all}a and Section \ref{sec:npdf}).

Over the entire $\nthp$ emission region, the long filament paths have much lower $N_{\rm tot}$ variation along their paths compared to the intersected ones. For example, Area-3 (Fig.\,\ref{fig:area_maps}, lower middle panel) contains three typical long paths. They have $N_{\rm tot}\leq 2\times 10^{23}$ cm$^{-2}$ over the total length of $\sim40''$. The $N_{\rm tot}$ values significantly increases only towards the ending points, where the paths are already intersected with other filaments. There are only a few cores located on the long filament paths, such as the one in Area-2 (OMC-3 MMS-7, labeled with a blue arrow). There are two candidate cores on this path, but the overall $N_{\rm tot}$ distribution still further increases towards the two ending points, which are both intersections.      

\subsection{Dense cores: Velocity Dispersion}
With a channel width of $\Delta V_{\rm chan}=0.1~\klms$, the $\nthp$ line profile can be resolved at a favorable accuracy, enabling us to measure the velocity dispersion of each core thereby estimate the internal pressure support against the self-gravity. 

% The velocity gradient over the different cores and filaments due to the gas motion will be analyzed in subsequent works. By fitting all the HFCs, we can measure the line width with an accuracy of 0.08 \kms, which exceeds the channel width of 0.1 \kms.   

The $\nthp$ (1-0) velocity dispersion is measured together with $T_{\rm ex}$ from the spectral fitting. Fig.~\ref{fig:nh2_profile}c and \ref{fig:nh2_profile}d exhibit the $\sigma_v(r)$ profiles along the major and minor axes of each local peak, respectively. Unlike the $N_{\rm tot}(r)$ profiles, $\sigma_v(r)$ profiles do not exhibit any evident increase towards the core centers. Instead, they are mostly confined in the range of $\sigma_v=(0.05,0.25)$ \kms\ over offset of $(-10'',10'')$. The average $\sigma_v(r)$ profiles even exhibit a slight decrease from $\sigma_v=0.13$ to 0.10 \kms\ towards the center along both the major and minor axes. This result suggests that the candidate dense cores tend to have comparable or even smaller velocity dispersion than the surrounding gas.

% thermal motion: velocity ---> temperature 
From the temperature and velocity dispersion, one can estimate the non-thermal turbulence scale using 
\begin{equation}\label{equ:cs_eff1}
\sigma_{\rm nt} = \sqrt{\sigma_v^2-\frac{k_{\rm B} T_{\rm kin}}{m_{\rm mol}}},
\end{equation}
where $m_{\rm mol}$ is the molecular mass, which is $m_{\rm mol}$=$m_{\nthp}$=$29m_{\rm H}$. Adopting a kinetic temperature of $T_{\rm kin}=T_{\rm ex}=15$ K, the observed $\sigma_v$ range leads to $\sigma_{\rm nt}$=0.03 to 0.24 \kms. Compared to the average sound speed of $c_s=\sqrt{k_{\rm B} T_{\rm kin}/\mu m_{\rm H}}=0.23~\klms$, the filaments tend to have an overall subsonic but non-zero level of turbulence, i.e., $0<\sigma_{\rm nt}/c_s\leq1$. And the slightly decreased $\sigma_v$ towards the center would likely trace the turbulence decay during the dense core formation, which is comparable to the observed trend in infrared dark clouds \citep[e.g.][]{wang08}.

\section{Analysis and Discussion}\label{sec:discussion}
\subsection{The Column-density Probability Distribution Function}\label{sec:npdf}
As introduced in Sec.\,\ref{sec:intro}, the N-PDF, especially its high-density tail, could monitor the gas component with likely turbulence decay and increased self-gravity. To reveal the gas assembly around the intersections, we compared the N-PDF around the intersections and other areas on the filaments (denoted as filament paths).  Around the intersections, the N-PDF is sampled in a circular area with $r = 7''$. Along the filament paths, it is sampled in a strip with width of $w = \overline{w}_{\rm fil,full}=14''$. To make comparison, we also sampled the $N_{\rm tot}$ distribution in the observed region out of the ISF, which contains the isolated gas patches and diffuse gas. Their $N_{\rm tot}$ distributions are shown in Fig.\,\ref{fig:npdf_all}. 

As shown in Fig.\,\ref{fig:npdf_all}a, the two components and outer region all have nearly gaussian-shaped $N_{\rm tot}$ distributions. The outer region is peaked at nearly zero level and declines to half maximum at $N_{\rm tot}=0.4\times 10^{23}$ cm$^{-2}$, slightly higher than the $5\sigma$ detection limit. This is consistent with its major component of weak diffuse gas around or bellow the detection limit. The second component (filament paths) is peaked at $N_{\rm tot}$=$0.4\times 10^{23}$ cm$^{-2}$ and is clearly separated from the component of outer region. The third component (intersections) is distributed towards further higher values, peaked at $N_{\rm tot}$=$1.4\times 10^{23}$ cm$^{-2}$. From their $N_{\rm tot}$ distributions, we can see that the diffuse gas components would not significantly affect the $N_{\rm tot}$-statistics of the filament paths and intersections. 

% In the range of $N_{\rm tot}>1.4\times 10^{23}$, the intersections have a pixel counting higher than the filaments by a factor of $\sim4$. 

% N-PDF, especially the high-end is insensitive to r_intsec
% We calculate whole the $N(H_{tot})$ of filament and cloud are $1.9\times10^{23}$ and $ 4\times10^{28}$, respectively. The filament is around 50\% of the cloud.
% Usually the local emission peaks near the intersections is covered by the sampling area, and the 

The $N_{\rm tot}$ distributions in the logarithmic scale, which are normally adopted to sample N-PDF, are presented in Fig.\,\ref{fig:npdf_all}b. The N-PDF would usually contain a lognormal component in lower $N_{\rm tot}$ range and a power-law tail towards the higher values \citep{kainulainen09}, with the second one possibly tracing the gas assembled due to the self gravity. The transition from lognormal to power-law components occurs around $\log (N_{\rm tot}/\sc)$=$23.3$ to 23.5, corresponding to $N_{\rm tot}$=$(2.0-3.2)\times 10^{23}$ cm$^{-2}$. At $\log N_{\rm tot}>23.3$, the N-PDF profiles of the both components switch into steeper slopes that can be fitted by a power-law shape of ${\rm d}N_{\rm count}/{\rm d}(\log N_{\rm tot}) \propto N_{\rm tot}^{\alpha}$. The best-fit indices are $\alpha$=$-2.8\pm0.2$ and $-2.5\pm0.1$ for the filaments and intersections, respectively. 

Compared to the intersections, the filament paths have more drastic decline over the transition point, their N-PDF drops to much lower level than the intersections when $\log N_{\rm tot}>23.3$. Towards higher $N_{\rm tot}$ values, the intersections have a pixel counting higher than the filaments by a factor of 4. The intersections have not only higher $N_{\rm tot}$ but also a bump-like feature above the average power-law trend in the range of $\log N_{\rm tot}=23.7$ to 24.0. This range is comparable to the peak values in the $N_{\rm tot}(r)$ profiles (Fig.\,\ref{fig:nh2_profile}, upper panels), suggesting that the dense cores should have a major contribution to the $N_{\rm tot}$ bump. 

% The power-law interval is determined after a large number of test. For some positions, The surrounding PDF has an evident power-law tail, but declines very rapidly, e.g. almost down to zero at logN = 23.3. If the lower-limit of the power-law tail is larger than 23.3, these positions would be excluded, the spatial sampling of the PDF would thus have a bias. On the other hand, the peak value of NPDF is always smaller than 23.2, it should thus be reasonable to adopt a fixed lower limit of logN = 23.2, in order to estimate the relative flatness of the NPDF tail at different positions.

% key points:
% local-maximum column density accompanied by declined velocity dispersion. 
% IR-cores may have additional support to the parental core due to heating and Ek output.
% if Pic is very large, the sub-virial cores (Mc<Mvir) can also collapse. 

\subsection{Inspecting the Spatial Variation of the N-PDF}
In addition to the global N-PDF, we also sampled the N-PDF over the different areas to inspect whether its variation has a real connection to the intersections or is merely due to the random fluctuation. 

First, we calculated the N-PDF in each sub-area, as shown in Fig.\,\ref{fig:npdf_all}c and Fig.\,\ref{fig:npdf_all}d, wherein intersections and filament paths are also separately measured. As a result, among all the sub-areas, the intersections have higher $\alpha$ values than the filament paths, suggesting that the more flattened N-PDF tails around the intersections should be a common trend. 

As shown in Fig.\,\ref{fig:npdf_all}c and \ref{fig:npdf_all}d, Area-3 and 5 have not only higher $\alpha$ than the other three sub-areas, but also highest $N_{\rm tot}$ that extend to $10^{24}$ cm$^{-2}$. One can see that Area-3 contains the most massive clump OMC-2 FIR-4 in Orion A north, which has a total mass of $39$ to $80~\msun$ as measured with different tracers and instruments \citep{crimier09,lopez13,nutter07,sadavoy10}. Area-5 also has a clump with considerable mass of $3$-$4~\msun$ \citep{nutter07,sadavoy10}. Although the other sub-areas also have relatively high total masses ($\geq 6\,\msun$), their overall gas distributions appear more extended and do not contain such dense and compact parental clumps over the filament structures. The comparison suggests that $\alpha$ would increase in the gas structures located in compact and massive clumps.

As a more detailed inspection, we estimated the N-PDF distribution over all the filament paths. We sampled the N-PDF along the filament paths in a step of $1.3$ arcsec (nearly half beam size). At each point, the N-PDF is measured in a circular area with radius of 14 arcsec. A number of selected N-PDF profiles over OMC-2,3 region are presented in Fig.\,\ref{fig:npdf_map}a. These N-PDF profiles all evidently exhibit a turn-over to the power-law tail at the column density range of $\log N_{\rm tot}>23.2$. 

Fig.\,\ref{fig:npdf_map}b and \ref{fig:npdf_map}c show the $\overline{N}_{\rm tot}$ and $\alpha$ distributions over the filaments in OMC-3, respectively. We note that the N-PDF distribution over entire OMC-2,3 region will be presented in subsequent works. As Fig.\,\ref{fig:npdf_map} shows, $\overline{N}_{\rm tot}$ and $\alpha$ distributions exhibit comparable variation trends. In particular, the area of MMS-2,3 has the highest values in both $\overline{N}_{\rm tot}$ and $\alpha$.   

For each point on the filament paths with N-PDF sampling, we measured its distance from the nearby intersection, denoted as $d_i$. The $\overline{N}_{\rm tot}$ and $\alpha$ distributions as functions of $d_i$ are shown in Fig.\,\ref{fig:npdf_stat}a and \ref{fig:npdf_stat}b, respectively. The two parameters both have much higher values around $d_i=0$ and a rapid decline towards larger $d_i$. In Fig.\,\ref{fig:npdf_stat}a, the gas component with $\overline{N}_{\rm tot}>2.0\times 10^{23}$ cm$^{-2}$ is confined within $d_i\leq 12''$ and become largely absent beyond this limit. The average level (black squares with error bars) exhibits a smooth decline with $d_i$. In Fig.\,\ref{fig:npdf_stat}b, $\alpha$ decreases from $\alpha=-2.4$ to $-3.3$ over $d_i=0$ to 15 arcsec, and then rises again towards $d_i=25$ arcsec. The rising feature at the high end could be due to the several individual dense cores away from the intersections. Based on these decreasing trends with $d_i$, the intersections should have a significant trend of assembling dense gas. From Fig.\,\ref{fig:npdf_stat}, one can also see that the gas component with high-$N_{\rm tot}$ also tends to have higher $\alpha$. According to the theoretical N-PDF analysis \citep{krumholz05,hennebelle11,padoan11}, it suggests that the gas therein could be more intensely bounded by the self-gravity. 

However, as shown in Fig.\,\ref{fig:npdf_stat}a, around the intersections there are also a large amount of gas with relatively low column density, i.e., $\overline{N}_{\rm tot}<2.0\times 10^{23}$ cm$^{-2}$. This amount of gas has low power-law slope of $\alpha=-6.5$ to $-2.0$ as shown in Fig.\,\ref{fig:npdf_stat}b. It suggests that some intersections do not have largely increased ${N}_{\rm tot}$. In other words, the intersected filaments should be a necessary but not sufficient condition for assembling the dense cores.

\subsection{The Gravitational Instabilities}\label{sec:m_vir}
Due to their low turbulence (Fig.\,\ref{fig:nh2_profile}), the candidate cores should be possibly bounded by the self-gravity. We calculated their virial masses, which can monitor the binding state due to the self gravity. For an ellipsoidal core, the virial mass can be calculated following \citet{li13}:
\begin{equation}\label{equ:m_vir}
%M_{\rm vir} = 210 \frac{R}{{\rm pc}} \frac{\Delta V}{\klms}~\msun,
M_{\rm vir} = \frac{5}{\alpha \beta} \frac{\overline{r}\, c_{\rm s,eff}^2}{G},
\end{equation}
where $\overline{r}$ is the average core radius, $\alpha$=$(1-a/3)/(1-2a/5)$ is a correction factor for the power-law density profile of $\rho \propto r^{-a}$. Using the value of $a=1.6$ for an isothermal cloud in equilibrium, we would have $\alpha=1.3$. $\beta = \arcsin e/e $ is the geometry factor to account for the eccentricity $e=\sqrt{1-f^{2}}$, where $f$ is the intrinsic ratio between minor and major axes. It can be estimated from the observed value using $f=(2/\pi) f_{\rm obs}F_{1} (0.5,0.5,-0.5, 1.5,1,1-f_{\rm obs}^{2})$, where $F_{1}$ is the first-kind Appell hypergeometric function, and $f_{\rm obs}$=$(r_{\rm min}/r_{\rm maj})_{\rm obs}$. Based on Eq.\,\ref{equ:cs_eff1}, the effective sound speed $c_{\rm s,eff}$ can be estimated using
\begin{equation}\label{equ:cs_eff2}
\begin{aligned}
c_{\rm s,eff} & =\sqrt{c_{\rm s}^2+\sigma_{\rm nt}^2} \\
              & =\sqrt{\sigma_{\rm obs}^2+ k_{\rm B} T_{\rm kin}\left(\frac{1}{\mu m_{\rm H}} - \frac{1}{m_{\rm mol}}\right)}.
\end{aligned}              
\end{equation}
If the core mass satisfies $M_{\rm core}\geq M_{\rm vir}$, i.e., $\alpha_{\rm vir}=M_{\rm vir}/M_{\rm core}\leq 1.0$, the thermal-and-kinetic energy should be lower than the potential well of the self-gravity so that the core would tend to be constrained by the self-gravity.

Another criterion to evaluate the core stability is the Bonner-Ebert (BE) mass, which would more specifically represent the mass upper-limit for a stable core because it considers two additional physical conditions, density gradient throughout the core and the external pressure. These two factors are important for constraining the gas in a dense core. The BE mass is estimated following the classical method \citep{stahler05} as:
\begin{equation}\label{equ:m_be}
M_{\rm BE}=\frac{m_1 c_{\rm s,eff}^4}{P_{\rm ic}^\frac{1}{2}G^\frac{3}{2}}
\end{equation}
wherein $m_1$=$1.18$, $P_{\rm ic}$=$n_{\rm ic} \mu m_{\rm H} \sigma_{\rm ic}$ is the external pressure onto the dense core, which depends on the external gas number density $n_{\rm ic}$ and velocity dispersion $\sigma_{\rm ic}$. The $\nht$ (1,1) and (2,2) observation with the Green Bank Telescope (GBT) \citep{kirk17} reveals an average pressure of $P_c/k_{\rm B}=2\times10^7$ K $\cc$ in the Orion A dense cores with a small variation. The single-dish $\nht$ cores sampled at a resolution of $10''$ should be comparable to the spatial sizes of the parental gas clumps of the filaments observed in the current work. It should be thus reasonable to adopt an approximation of $P_{\rm ic}=P_c$ in our calculation.

Following \citet{li13}, we also estimated the mass fraction that can be supported by the magnetic field $B$,
\begin{equation}\label{equ:m_phi}
M_{\rm \Phi}=c_{\rm \Phi}\frac{\pi\, B \overline{r}^2}{G^{1/2}},
\end{equation}
wherein the coefficient is $c_{\rm \Phi}$=$0.12$, and the magnetic field in OMC-2,3 could vary from $B$=0.64 to 0.85 mG \citep{houde04,matthews05}, causing a mass range of $\overline{M}_{\rm \Phi}=0.14$ to 0.19 $\msun$ at the average radius of $\overline{r}=5.3\times 10^{-3}$ pc. Combining the hydrostatic and magnetic components together, we have the critical mass of
\begin{equation}\label{equ:m_crit}
M_{\rm crit} = M_{\rm BE} + M_{\rm \Phi},
\end{equation}
The physical properties of the $\nthp$ cores and the derived $M_{\rm crit}$ and $M_{\rm vir}$ are presented in four diagrams in Fig.\,\ref{fig:r_mvir}. 

Fig.\,\ref{fig:r_mvir}a shows the $c_{\rm s,eff}$ and $M_{\rm core}$ distributions. The $\nthp$ cores have $c_{\rm s,eff}=0.25$ to 0.5 \kms and $M_{\rm core}$=0.07 to $1.8~\msun$. The major fraction of the cores have $c_{\rm s,eff}<0.35~\klms$. Although several cores with high masses tend to have slightly increased $c_{\rm s,eff}$, the overall sample does not have noticeable correlation between $c_{\rm s,eff}$ and $M_{\rm core}$. It suggests that low turbulence should be a universal state for all the cores in spite of their masses. 

Fig.\,\ref{fig:r_mvir}a also shows the 850 $\micron$ cores observed with JCMT/SCUBA-2 \citep{kirk17}, wherein the cores associated with the $\nthp$ filaments are denoted with the black-edged diamonds. The velocity dispersion of the 850 $\micron$ cores were measured from the GBT $\nht$ lines. The major fraction of these cores have $M_{\rm core}$=$0.18$ to $3.0\,\msun$, while the parental cores of the $\nthp$ filaments (black-edged diamonds) tend to have relatively high masses from 0.7 to 10 $\msun$. It suggests the relatively high-mass cores are located within the ISF. The three samples of cores also have distinct $c_{\rm s,eff}$ distributions. The parental cores have lower average $c_{\rm s,eff}$ than the other 850 $\micron$ cores. And the $\nthp$ cores have further lower $c_{\rm s,eff}$ than their parental cores. This comparison suggests a decline trend for the turbulence from large to small scales, in particular towards the inner regions of the large-scale cores.  

Fig.\,\ref{fig:r_mvir}b shows the $M_{\rm core}$ and $r_{\rm core}$ distribution of the $\nthp$ cores. It shows that $M_{\rm core}$ is increasing with $r_{\rm core}$. Their relation can be best fit with a power law of 
\begin{equation}
\frac{M_{\rm core}}{\msun} = 5.8\times 10^{-3}\left(\frac{r}{10^{-3}\,{\rm pc}}\right)^{2.9}.
\end{equation}
The power-law index (2.9) is close to 3.0, suggesting that the cores have similar number densities, so that the core mass can be approximated as $M_{\rm core}=\mu m_{\rm H} n_{\rm core} (4/3) \pi r^3$. The average number density is estimated to be $n=(1.8\pm0.6) \times 10^7$ $\cc$. The 850 $\micron$ cores have $r>0.017$ pc and their radii are not overlapped with the $\nthp$ cores (insert panel). This is due to the lower limit of $r_{\rm core}$ set by the JCMT beam size. From their $M(r)$ distribution, the 850 $\micron$ cores are estimated to have much lower densities of $n=(2.0\pm1.5) \times 10^6$ $\cc$. In comparison, the average density of the $\nthp$ cores ($1.8 \times 10^7$ cm$^{-3}$) would represent a characteristic value of the dense-gas assembly in the large-scale cores. As suggested by the difference in $c_{\rm s,eff}$ between the 850 $\micron$ and $\nthp$ cores, the gas assembly would be closely related to the turbulence dissipation. During the turbulence dissipation, the dense gas would gradually reach a hyndrostatistic equilibrium (BE sphere). In this condition, the cores would have a limited possibility for further compression, unless they collapse into protostars.

In Fig.\,\ref{fig:r_mvir}b, the theoretical functions of $M_{\rm vir}(r)$ and $M_{\rm crit}(r)$ are also plotted. They are calculated from Eq.\,\ref{equ:m_vir} and Eq.\,\ref{equ:m_crit}, respectively. In calculation, an average value of $c_{\rm s,eff}=0.3~\klms$ is adopted. As shown in the diagram, the cores mostly have $M_{\rm core}>M_{\rm vir}$ (super-virial), with only about ten cores having $M_{\rm core}<M_{\rm vir}$ (sub-virial). In comparison, a much more fraction of the cores have $M_{\rm core}<M_{\rm crit}$ (sub-critical). It is noteworthy that $M_{\rm crit}(r)$ has a minor but not negligible contribution from $M_{\rm \Phi}(r)$, namely the magnetic field. Because of the $M_{\rm \Phi}$-contribution, $M_{\rm crit}(r)$ rises above the $M_{\rm BE}$ level for a scale of $\sim0.3\,\msun$ towards large radius. As a result, about $\sim10$ cores are located between the two levels, namely $M_{\rm BE}<M_{\rm core}<M_{\rm crit}$. For these cores, the magnetic field may play a significant support against the self-gravity.          

% $M_{\rm crit,obs}>M_{\rm crit}(r)$, suggesting them to be possibly unstable to the self-gravity. 

To more specifically estimate the stability of the individual cores, we derived the virial parameter $\alpha_{\rm vir}=M_{\rm vir}/M_{\rm core}$ and critical mass ratio $\alpha_{\rm crit}=M_{\rm crit}/M_{\rm core}$. The $\alpha_{\rm vir}(M_{\rm core})$ and $\alpha_{\rm crit}(M_{\rm core})$ distributions are shown in Fig.\,\ref{fig:r_mvir}c and Fig.\,\ref{fig:r_mvir}d, respectively. Actually, these two diagrams exhibit similar results with Fig.\,\ref{fig:r_mvir}b except a slight difference in the number of cores above and bellow each threshold. As shown in the two panels, most cores (7 exceptions) have $\alpha_{\rm vir}<1$, suggesting them to be bounded by the self-gravity. A less fraction of the cores, about 30 (23 \%), have $\alpha_{\rm crit}<1$. This result indicates that cores are confined by the self-gravity, but only about one fifth of the cores would have the tendency to collapse into protostars. 

It is noted that there are two uncertainties in deriving the core instabilities. First, it is uncertain whether the magnetic support is comparable to that characterized by $M_{\Phi}$ and whether most cores have similar B-field with the previous measurement. As shown in Fig.\,\ref{fig:r_mvir}b, if only using $M_{\rm BE}$, the super critical cores would moderately increase by a number of $\sim10$, or a fraction increased to 30 \%.   

Second, the external pressure $P_{\rm ic}$ is also from a less direct estimation. The actual value may be lower ($P_{\rm ic}<P_{\rm c}$), because the single-dish cores \citep{kirk17} may not throughly and uniformly harbor the $\nthp$ filaments. Moreover, in a clump, the pressure would decline from its center to outside, so the confinement to the internal filaments by the thermal pressure could be less effective. To increase the instabilities, one may consider other factors such as the dynamical pressure due to the external gas motions. For example, the gas converging flows along the filaments with several \kms\ would compress the central region, providing a dynamical pressure comparable to $P_{\rm ic}$. Since the gas flows are frequently observed in Orion and other regions from large to small scales \citep[e.g.][]{peretto14,hacar17,yuan18,chen19a}, they may considerably affect the core instability. However, the external perturbation would also inject hot gas and kinetic energy into the cores, increasing the temperature and turbulence, letting them become even less bound. Considering that $M_{\rm BE}$ is much more sensitive to $c_{\rm s, eff}$ than $P_{\rm ic}$ (Eq.\,\ref{equ:m_be}), the external perturbation is more likely to have a negative effect to the binding state of the cores unless the input energy can be efficiently dissipated. Based on all these factors, at the current state, the cold dense cores in Orion A seem unlikely to have very extensive star-forming activities.        

% The $\nthp$ dense cores directly related with the star-forming and earliest phase.
% The dense-core formation efficiency is... but the star-forming efficiency may depends on the fraction of the core mass that can eventually fall into the stars, and the fraction of the cores to collapse. 

\subsection{The Efficiency of Star Formation}
From the integrated $\nthp$ emission and using Eq.\ref{equ:nthp_ncol}, we estimated a total gas mass of $350~\msun$ for the ISF in OMC-2,3 region. In comparison, the dense cores altogether have a total mass of $69~\msun$, taking up 20 \% of the total gas mass.  We thus suggests a core-formation efficiency (CFE) of $\sim$0.2 for the cold dense gas ($T\simeq 15$ K, $n>10^5$ $\cc$) in Orion A north. 

From the dust continuum observation in Orion B, \citet{konyves20} measured a CFE of 0.01 to 0.2 from the Herschel 70 to 500 $\micron$ images at a typical resolution of $18''$. In particular, the relatively high column-density regions ($A_V>7$ or $N_{\rm tot}>10^{22}$ cm$^{-2}$) exhibit a relatively high value of CFE$>0.1$, which is comparable to the value in OMC-2,3. The similarity of CFEs at large and small scales could be maintained by the self-similar structures in filamentary clouds. In the current study, the $\nthp$ filaments could more particularly trace the dense gas that has low turbulence and are more confined by self-gravity. It therefore exhibits higher CFE than the single-dish cores observed at larger scales, which contain more extended and turbulent gas. 

Compared to the CFE, the fraction of gas mass to eventually form stars is more uncertain. Currently there is no confirmed protostellar objects among our $\nthp$ cores since the cores are all displaced from the catalogued YSOs \citep{megeath12}. Among the cores, we would first consider the supercritical ones ($M>M_{\rm crit}$) as the star-forming candidates, which have a total mass of $\sim50~\msun$. Assuming a stellar-to-core mass ratio of $\epsilon=0.8$ \citep{beuther02,wang10}, we can estimate an upper limit of total stellar mass to be $40~\msun$. 

% ++ also compared to li13
% a fraction of the cores are pressure-bounded. The external pressure would be a major trigger for the core collapse. 
% elongated gas structures, intersections are not biased towards isolated peaks 

\section{Summary and Conclusion}
\label{sec:summary}
% although the filaments may help stabilize the cloud, dense gas are still continuously assembled at the intersections.
We investigated the cold dense filaments and cores in OMC-2,3 region in Orion A North using the ALMA $\nthp$ (1-0) line emission, in particular the isolated hyperfine component $JF_1=1_0$-$0_1$. The filament paths were extracted using the DisPerSE algorithm. From the comparative study between dense gas distribution and the filament structure, we explored the physical properties in four major aspects.

(1) The filament extraction reveals the spatial and velocity distributions of the intersected filaments. With the likely pseudo-ones excluded, there are more than 170 intersections for the filaments in the ISF over a spatial scale of 2 pc. They have single-velocity component with relatively narrow line width ($\Delta V \leq 0.6$ \kms), thus would represent the different filaments being merged together instead of merely overlapped in projection.

(2) Along the main body of integral-shaped filament (ISF), 128 candidate cores are identified from the local emission peaks above the $5\sigma$ detection limit. The cores have small velocity dispersion of $\sigma$=0.03 to 0.24 \kms and a mass distribution from 0.07 to 1.9 $\msun$. A large fraction of the cores (103) are located around the intersections with offset less than 5 arcsec, while only a small fractions (25) are located on the long filament paths or displaced from the ISF. The comparison shows that the intersections would play a significant role in assembling the local dense gas, while the fragmentation of individual filament path into the cores are less prominent in our observed spatial scale (0.005 to 0.01 pc). 

% The filament extraction using DisPerSE is not spatially biased to the local peaks, 

(3) The N-PDFs of the intersections and other part of filaments both have an evident power-law tail towards high column densities, i.e. $\log (N_{\rm tot}/{\rm cm^{-2}})$>23.3. But the areas with high $N_{\rm tot}$ are mainly associated the intersections, and the power-law index of the N-PDF tail also exhibits an increasing trend towards the intersections, in agreement with the dense gas assembly therein. 

% In particular, Area-3 exhibit a bump-like feature around $\log (N_{\rm tot}/\sc)=23.8$ in its N-PDF, which should be mainly contributed by the high-mass dense clump OMC-2 FIR-4.

% Among the sub-areas, the N-PDF in Area-3 (northern part of OMC-2) has the most flattened power-law tail as well as the tail extending to highest $N_{\rm tot}$. The difference in Area-3 could be largely due to the massive gas clump OMC-2 FIR-4. This comparison suggests that the gas assembly around the intersections are proceeding in all the areas despite that they have largely different morphologies, but in the meantime, would still be more enhanced by the massive parental gas clump.

(4) Most of the cores (7 exceptions) have virial parameter of $\alpha_{\rm vir}\leq 1$. They should represent a sample of cold dense prestellar cores in OMC-2,3. However, only 23 \% of the cores may have potential to collapse into protostars due to their low critical mass ratio ($\alpha_{\rm crit}\leq 1$), while the other cores could be more stabilized by the internal velocity dispersion and magnetic field. In general, the cold prestellar cores in OMC-2,3 may require further mass growth or other perturbations to more broadly initiate their star formation.

\section{Acknowledgement}
We thank the referee for very detailed comments that help improve the analysis. This work is supported by the National Natural Science Foundation of China No. 11988101, No. 11725313, No. 11403041, No. 11373038, No. 11373045, CAS International Partnership Program No. 114A11KYSB20160008, and the Young Researcher Grant of National Astronomical Observatories, Chinese Academy of Sciences. 

% \bibliography{/Users/rzy/latex/bib_all}

% Don't change these lines
\bsp	% typesetting comment
\setcounter{table}{0}

\begin{table*}
\label{tab:observation}
{\small
\centering
\begin{minipage}{120mm}
\caption{Field Center Coordinates for the Mosaic Mapping. }
\begin{tabular}{llll}
\hline
No.     &   RA                &   Dec             &  On-source time$^a$  \\
\quad   &   (hh:mm:ss)        &   (dd:mm:ss)      &  (12-m, 7-m) (min)   \\
\hline
\multicolumn{3}{l}{(OMC-3)} \\              
01  &    05:35:14.51          &     -04:59:32.8   &  17.48, 13.96   \\
02  &    05:35:16.44          &     -05:00:06.4   &  17.48, 13.96   \\
03  &    05:35:18.26          &     -05:00:33.6   &  17.48, 13.96   \\
04  &    05:35:20.25          &     -05:00:59.2   &  17.48, 13.96   \\
05  &    05:35:22.11          &     -05:01:28.0   &  17.48, 13.96   \\
06  &    05:35:23.73          &     -05:01:57.6   &  17.48, 13.96   \\
07  &    05:35:25.44          &     -05:02:30.4   &  17.48, 13.96   \\
08  &    05:35:26.02          &     -05:03:10.4   &  17.48, 13.96   \\
09  &    05:35:26.35          &     -05:03:51.2   &  17.48, 13.96   \\
10  &    05:35:26.30          &     -05:05:29.6   &  17.48, 13.96   \\
11  &    05:35:26.83          &     -05:04:44.0   &  17.48, 13.96   \\ 
\multicolumn{3}{l}{(OMC-2)} \\
12  &    05:35:23.51          &     -05:07:24.8   &  17.48, 13.96   \\
13  &    05:35:24.64          &     -05:08:07.2   &  17.48, 13.96   \\
14  &    05:35:25.07          &     -05:08:54.4   &  17.48, 13.96   \\
15  &    05:35:26.30          &     -05:09:49.6   &  17.48, 13.96   \\
16  &    05:35:26.67          &     -05:10:36.0   &  17.48, 13.96   \\
17  &    05:35:25.12          &     -05:12:16.0   &  17.48, 13.96   \\
18  &    05:35:22.33          &     -05:12:36.8   &  17.48, 13.96   \\
19  &    05:35:20.83          &     -05:13:11.2   &  17.48, 13.96   \\
20  &    05:35:21.37          &     -05:14:20.8   &  17.48, 13.96   \\
21  &    05:35:20.46          &     -05:14:57.6   &  17.48, 13.96   \\
22  &    05:35:19.12          &     -05:15:31.2   &  17.48, 13.96   \\
23  &    05:35:18.42          &     -05:13:03.2   &  17.48, 13.96   \\
24  &    05:35:22.44          &     -05:10:12.0   &  17.48, 13.96   \\
25  &    05:35:25.81          &     -05:11:26.4   &  17.48, 13.96   \\  
\hline
\end{tabular} \\
$a.${ The On-source integration time for each pointing (see Fig. 1 for \\
the mapping areas of the pointings). The two values correspond to 12-meter \\
and 7-meter (ACA) arrays, respectively. }  \\
\end{minipage}
}
\end{table*}

% ------------ Table 2 ----------------
\begin{table*}
\label{tab:parameters}
{\small
\centering
\begin{minipage}{120mm}
\caption{The physical properties of the gas structures. }
\begin{tabular}{llll}
\hline\hline
Parameters                    &   Cores                           &   Filament paths${\rm ^b}$     &   Unit     \\
\hline                                                                                                            
$2 R_{\rm maj}$ $^{\rm a}$    &   $8.5\pm2.0$, $3.3\pm 0.7$       &   $-$                          &   arcsec, $10^3$ AU         \\
$2 R_{\rm min}$ $^{\rm a}$    &   $7.5\pm2.0$, $2.9\pm 0.3$       &   $-$                          &   arcsec, $10^3$ AU         \\
Length                        &   $-$                             &   5-60, 1.9-23                 &   arcsec, $10^3$ AU         \\
Width                         &   $-$                             &   $14\pm 4$, $5.0\pm 1.5$      &   arcsec, $10^3$ AU         \\
Mass                          &   $1.7\pm 0.5$                    &   $2.0\pm 0.5$                 &   $\msun$                   \\
$\sigma_{\rm v}$ $^{\rm c}$   &   $0.12\pm 0.05$                  &   $0.15\pm 0.07$               &   \kms                     \\                   
$T_{\rm ex}$ $^{\rm d}$       &   $15\pm 3$                       &   $15\pm 4$                    &   K                         \\
$N_{\rm tot}$ $^{\rm e}$      &   $4.0\pm 2.5$                    &   $2.0\pm 0.5$                 &   $10^{23}$ cm$^{-2}$       \\
\hline
\end{tabular} \\
a.{ The average FWHM diameter deconvolved with the synthesized beam size of 3 arcsec $\times$ 3 arcsec. }  \\
b.{ For the filaments, the values represent the distribution of the filament paths. Each individual path represents a segment between two endpoints, and an endpoint could be either an intersection or a terminal.} \\
c.{ The velocity dispersion inferred from the observed line width as $\sigma_{\rm v}=\Delta V / \sqrt{8 \ln 2}$. }  \\
d.{ The $\nthp$ excitation temperature estimated from the spectral fitting. }  \\
e.{ Derived from the integrated $\nthp$ emission using Equation \ref{equ:nthp_ncol} and assuming $\nthp$ abundance of $X(\nthp) = 7.5\times 10^{-10}$. }  \\
\end{minipage}
}
\end{table*}

% \clearpage
% \section{Figures}
\setcounter{figure}{0}
\begin{figure*}
	\centering
	\includegraphics[width=0.8\textwidth]{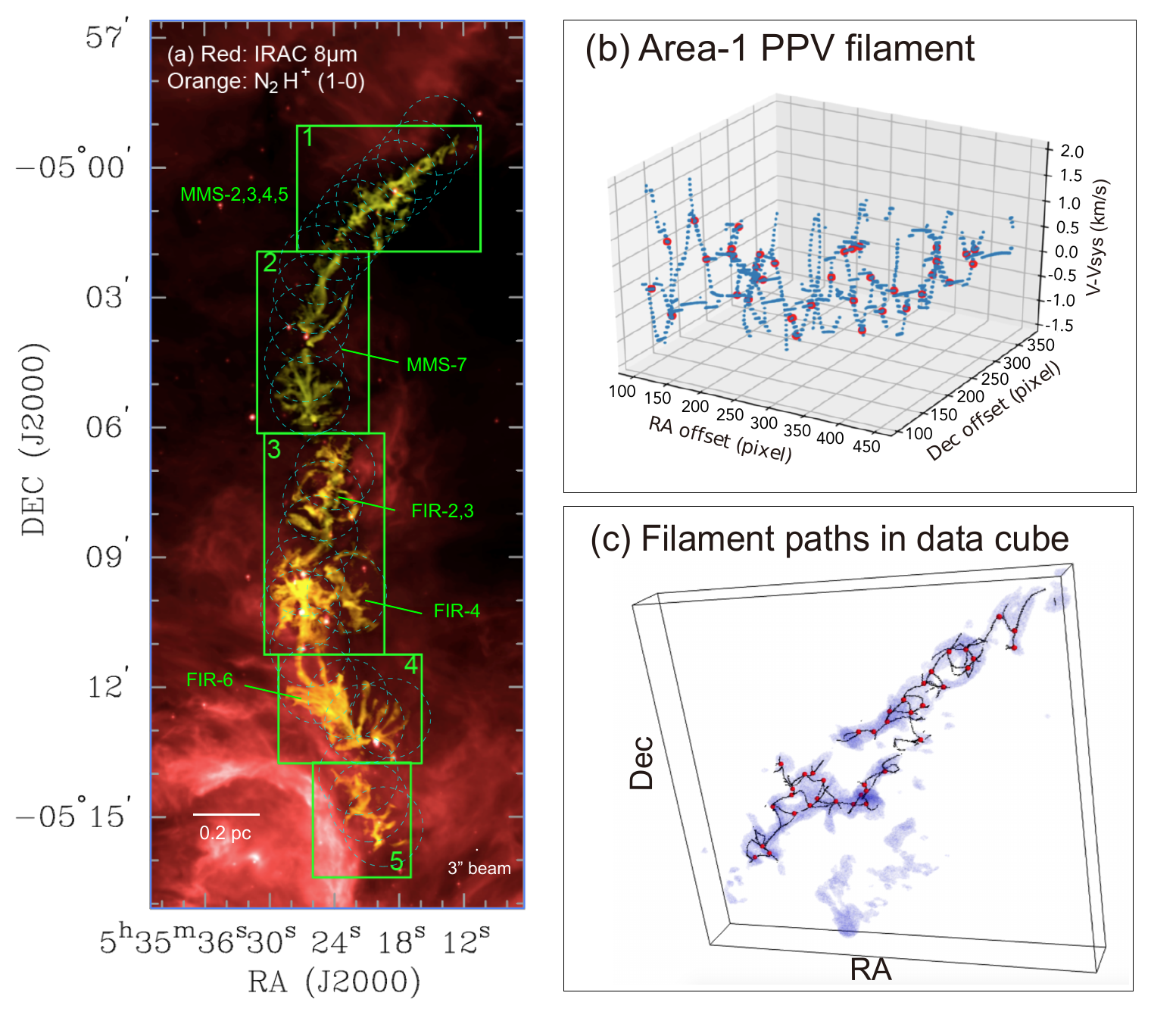}
	\caption{{\bf (a)} The integrated ALMA $\nthp$ $(1_0-0_1)$ image (yellow region) overlaid on the Spitzer/IRAC 8 $\micron$ image (red image). The $\nthp$ image only exhibits the central integral-shaped filament (ISF). The five sub-areas are labeled with boxes. Each area contains relatively condensed filament structures that are reasonably separated from the adjacent areas. Area-1,2,3 belong to OMC-3 and Area-4,5 belong to OMC-2. The dashed circles indicate the field of view (primary beam) of each pointing center for the mosaic mapping. {\bf (b)} An example of the filament paths modeled by DisPerSE in Area-1 in PPV space. The blue dashed lines represent the filament paths and the red dots are the intersections. {\bf (c)} The filament paths overlaid with the emission regions in the data cube. The blue-colored body represents the line emission in the data cube above the $5\sigma$ detection limit. }                                          
	\label{fig:omc23_ppv}
\end{figure*}

\begin{figure*}
	\centering
	\includegraphics[width=1.0\textwidth]{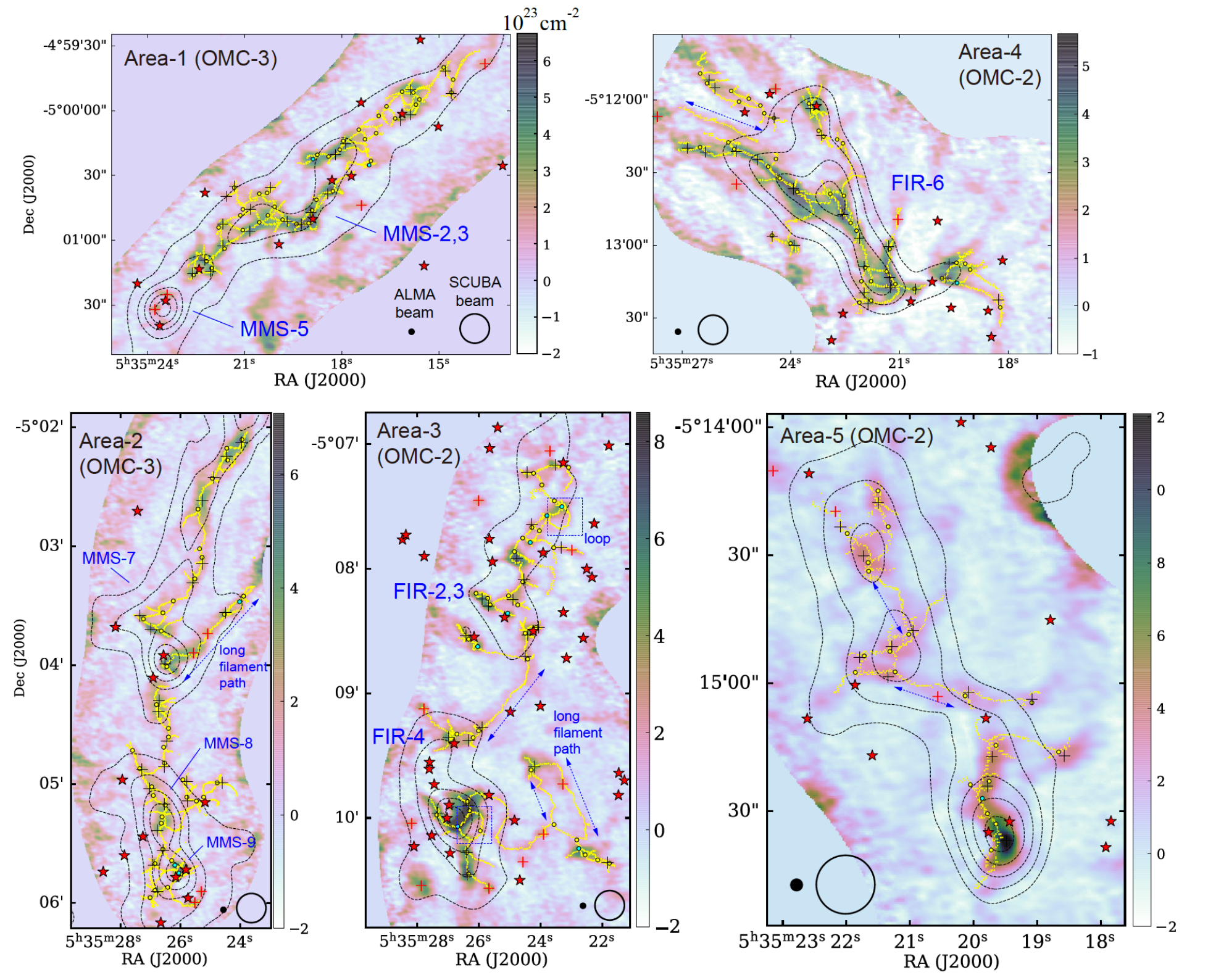}
	\caption{ The $\nthp$ emission in each sub-area (false-color image). The intensity is converted to the total gas column density in unit of $10^{23}$ cm$^{-2}$ using Eq.\,\ref{equ:nthp_ncol}. The overlaid dashed contours represent the JCMT/SCUBA 850 $\micron$ continuum map \citep{johnstone99}. The contour levels are 10, 30, 50, 70, 90 percent of the local maximum. Also overlaid on the figure are the DisPerSE-modeled filament paths (yellow dotted lines) projected onto the RA-Dec plane, real and pseudo intersections (yellow and cyan dots, respectively), local emission peaks (plus), and YSOs (stars) identified from infrared point sources \citep{megeath12}. Among the local emission peaks, the black ones denote those near the intersections within a distance of 7 arcsec ($\overline{w}_{\rm fil,FWHM}$), while the red ones denote those more distant from the intersections, e.g. those on the long filament paths or displaced from the ISF. The representative long paths and a typical encircled structure (loop) are labeled with blue dashed arrows and boxes, respectively. In each sub-area, The gas clumps and young stellar clusters are labeled with the blue characters following the denotation in \citet{li13} and references therein. }                                          
	\label{fig:area_maps}
\end{figure*}

\begin{figure*}
	\centering
	\includegraphics[width=0.7\textwidth]{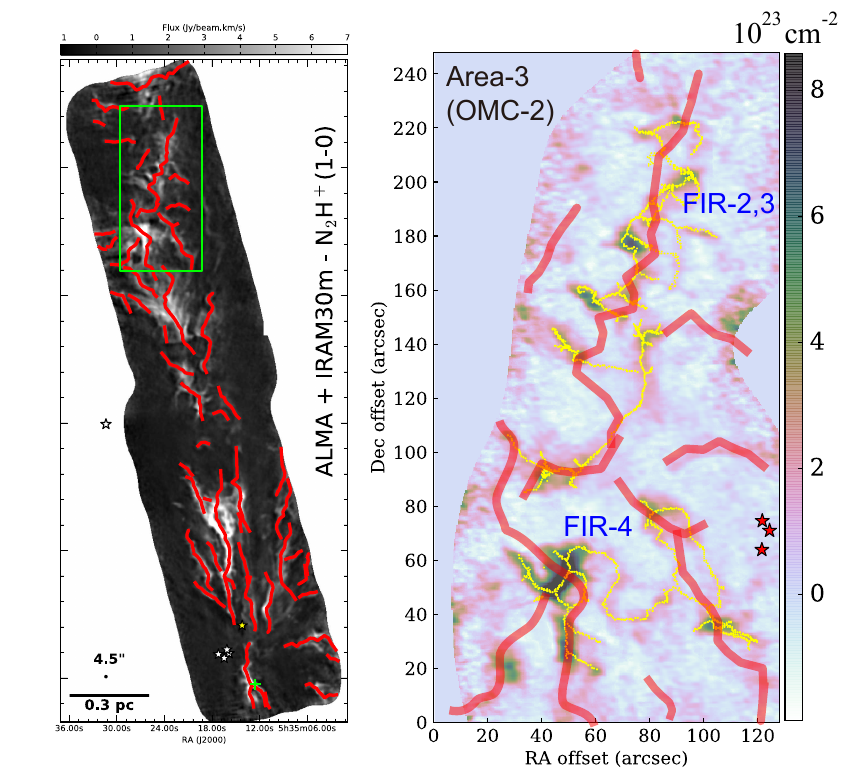}
	\caption{{\bf Left:} The principal axes of the fibers identified by the HiFIVE analysis over OMC-1,2,3 regions \citep[][Figure 4 therein]{hacar18}. {\bf Right:} The fibers (thick red segments) overlaid on the current $\nthp$ emission and the filament structures extracted by DisPerSE, same as that shown in Fig.\,\ref{fig:area_maps}. }                                          
	\label{fig:fila_compare}
\end{figure*}

\begin{figure*}
	\centering
	\includegraphics[width=0.9\textwidth]{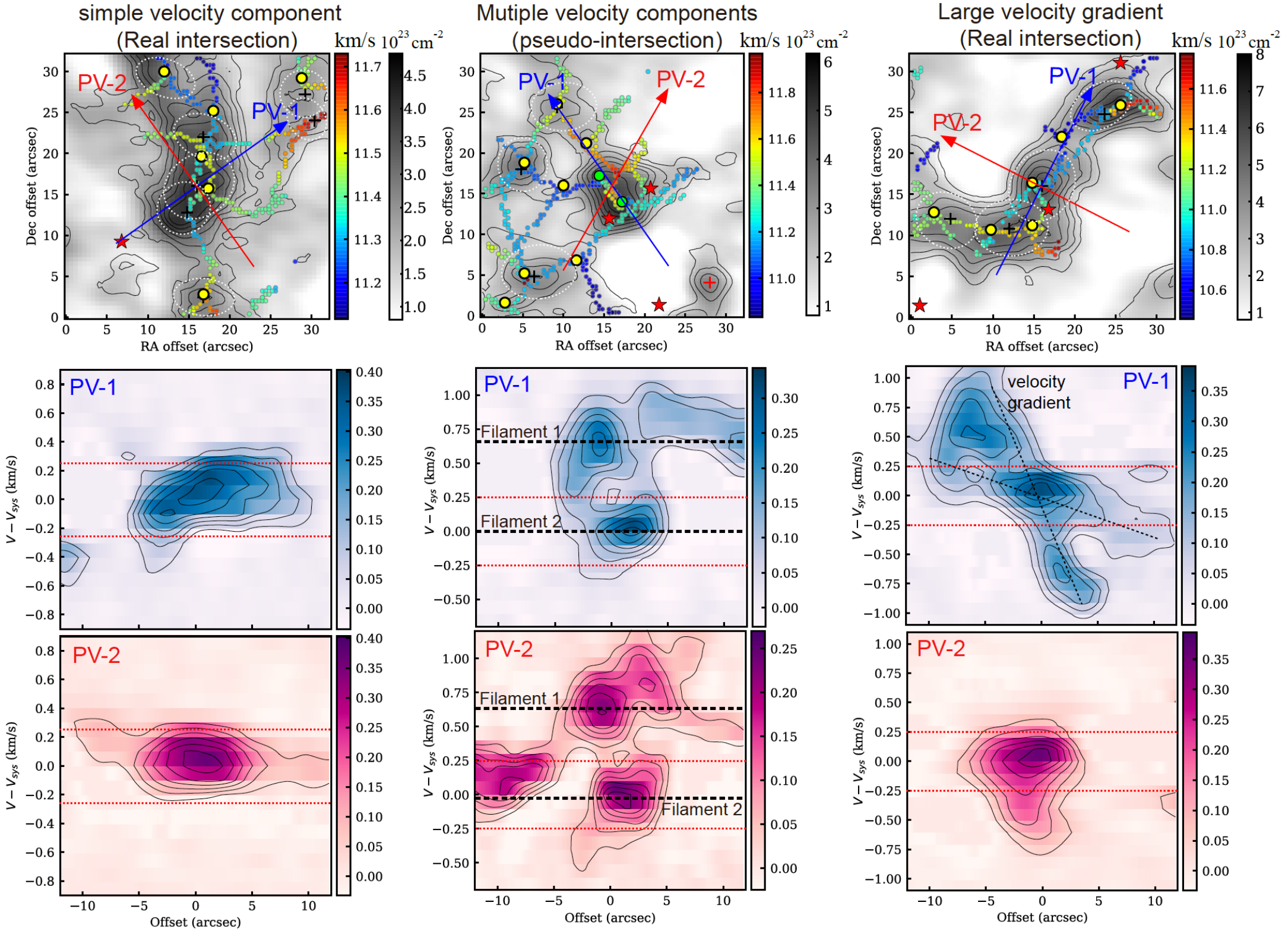}
	\caption{The gas distributions and Position-velocity diagrams over three representative gas structures, (i) real intersection with simple velocity component (in MMS-8,9), (ii) likely pseudo-intersection (in MMS-8,9), and (iii) real intersection with evident velocity gradient (in MMS-2,3). They are shown in the three columns, respectively. In each column, {\bf upper panel} is the integrated $\nthp~(1_0-0_1)$ image (gray scale and contours). The gray scale starts from $0.8\times 10^{23}$ cm$^{-2}$, which is the average level of the diffuse emission away from the ISF. The contour levels are 15, 30, 45, 60, 75, 90 percent of the local maximum. The colored dots represent the filament structures modeled by DisPerSE, with the color scale indicating the radial velocity. The yellow and green circles represent real and pseudo intersections, respectively. The crosses are the local emission peaks above $5\sigma$ level. The red stars are the YSOs from the catalogue \citep{megeath12}. The directions to plot the P-V diagrams are denoted with red and blue arrows. The two directions are along the major and minor axes of the central core. {\bf Middle and lower panels} are the P-V diagrams along the major and minor directions, respectively. The velocity components and velocity gradients are denoted with black dashed lines. The red dotted lines represent the velocity range of $\pm 0.25$ \kms corresponding to the average line width of 0.5 \kms.}                                          
	\label{fig:section_example}
\end{figure*}

\begin{figure*}
	\centering
	\includegraphics[width=0.5\textwidth]{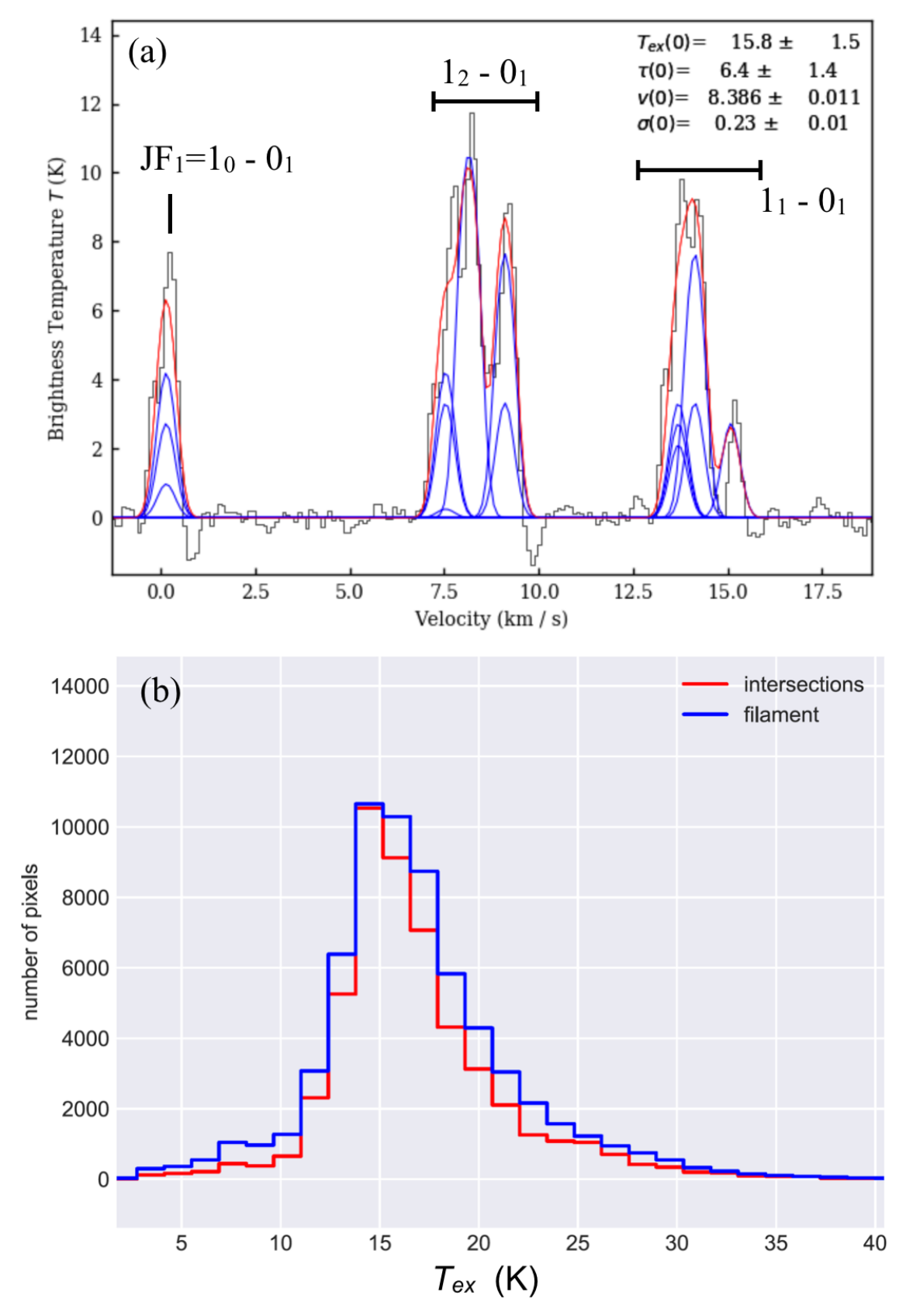}
	\caption{{\bf (a)} An example spectrum from OMC-2 FIR-6 and the hyper-fine fitting. The black line is the observed spectrum. The red line is the optimized model spectrum, with the individual HFCs shown in blue lines. Each HFC is a transition between $JF_1 F$ levels. The blue lines represent all the $JF_1 F$ transitions, and their frequencies are adopted from \citet{pagani09}. The three major groups of $JF_1$ transitions are labeled on the spectrum. We note that the $JF_1=1_0-0_1$ group contains three $F$ transitions with the same frequencies. The physical parameters of the best-fit spectrum are labeled on the panel. {\bf (b)} The histogram of $T_{\rm ex}$ distribution over the ISF. The areas around intersections and the filament paths are separately sampled. Their distributions are presented in red and blue histograms, respectively. }                                          
	\label{fig:spectra_fitting}
\end{figure*}

\begin{figure*}
	\centering
	\includegraphics[width=0.7\textwidth]{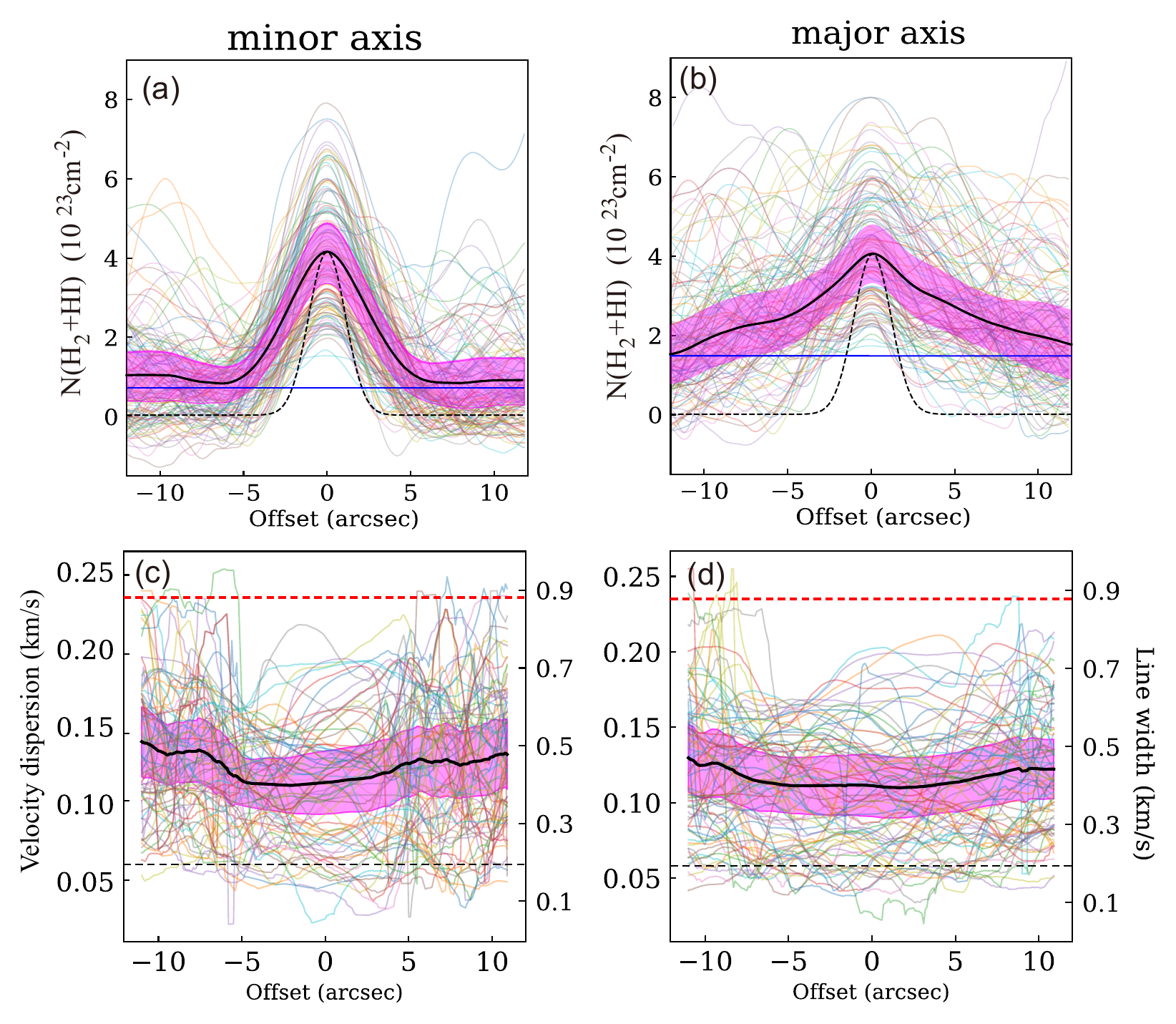}
	\caption{{\bf Upper panels:} The column density profiles $N_{\rm tot}(r)$ over the dense cores (local emission peaks). Each profile is centered at the local emission peak. The thick solid black line represents the average profile for all the peaks. The light purple strip represents the standard deviation around the average level. The horizontal solid line indicates the average surrounding level of $N_{\rm tot}$ away from the cores. The dashed line represents the synthesized beam shape. The left and right panels are sampled along the minor and major axes of each peak, respectively. {\bf Lower panels:} The velocity dispersion (line width) profiles of the $\nthp$ $(1_0-0_1)$ line over the cores. In each panel, the solid line and strip also represent the average level and standard deviation, respectively. The red and black dashed lines represent the thermal motion at $T_{\rm kin}=15$ K for the $\htwo$ and $\nthp$ molecules, respectively.}                                          
	\label{fig:nh2_profile}
\end{figure*}

\begin{figure*}
	\centering
	\includegraphics[width=0.8\textwidth]{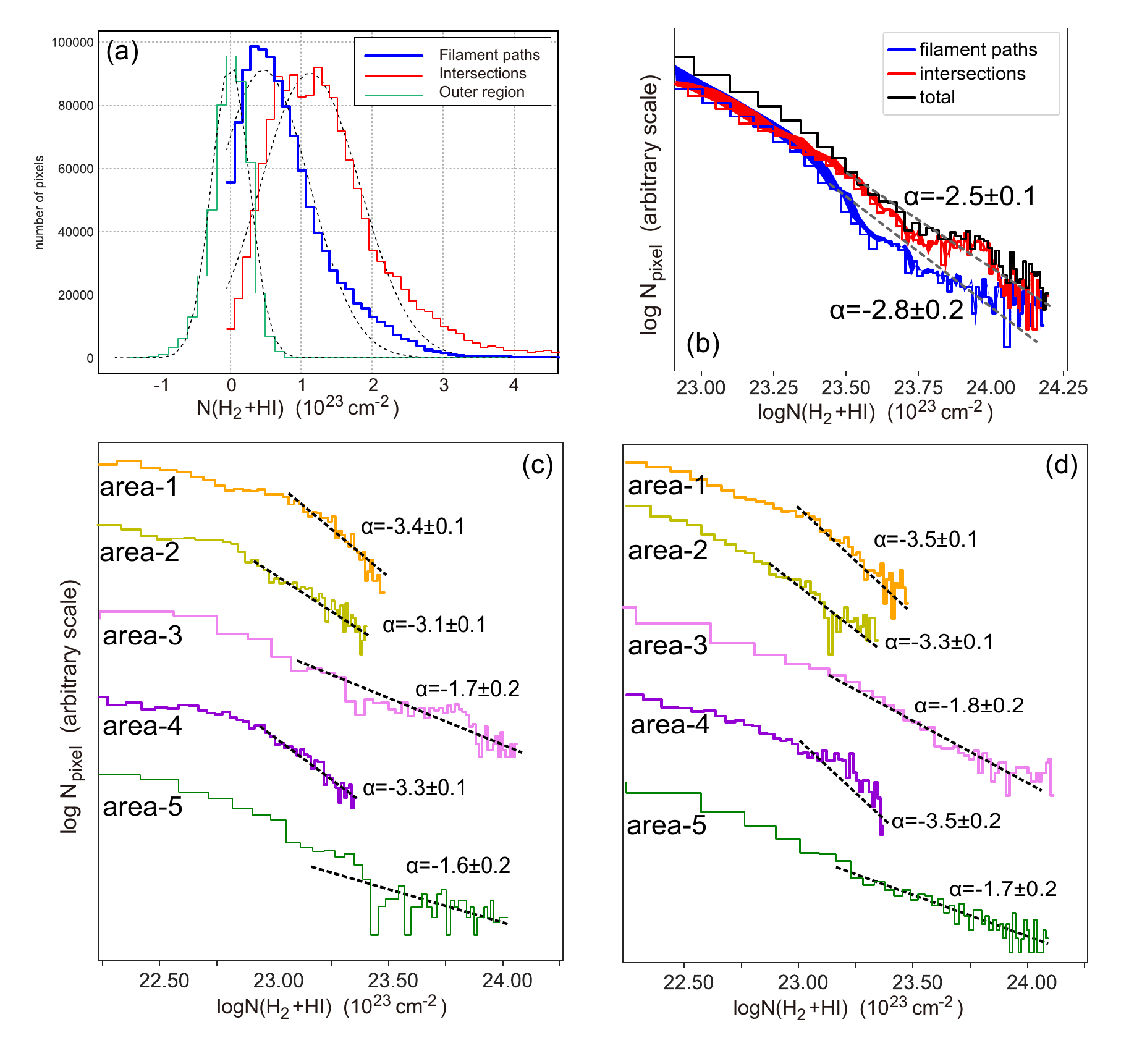}
	\caption{{\bf (a)} The $N_{\rm tot}$ distributions of the three components in linear scale: intersections (red), filament paths (blue), and the outer region away from the ISF (green). The black dashed lines represent the gaussian fitting for each component. {\bf (b)} the $N_{\rm tot}$ probability distribution function (N-PDF) of the intersections (red line) and filaments (blue line) in logarithmic scale. The black line is the summation of the two components, namely the total N-PDF of the ISF. {\bf (c)-(d)} The N-PDFs sampled in the sub-areas for {\bf (c)} intersections and {\bf (d)} the filament paths with the intersections excluded.}                                          
	\label{fig:npdf_all}
\end{figure*}

\begin{figure*}
	\centering
	\includegraphics[width=0.7\textwidth]{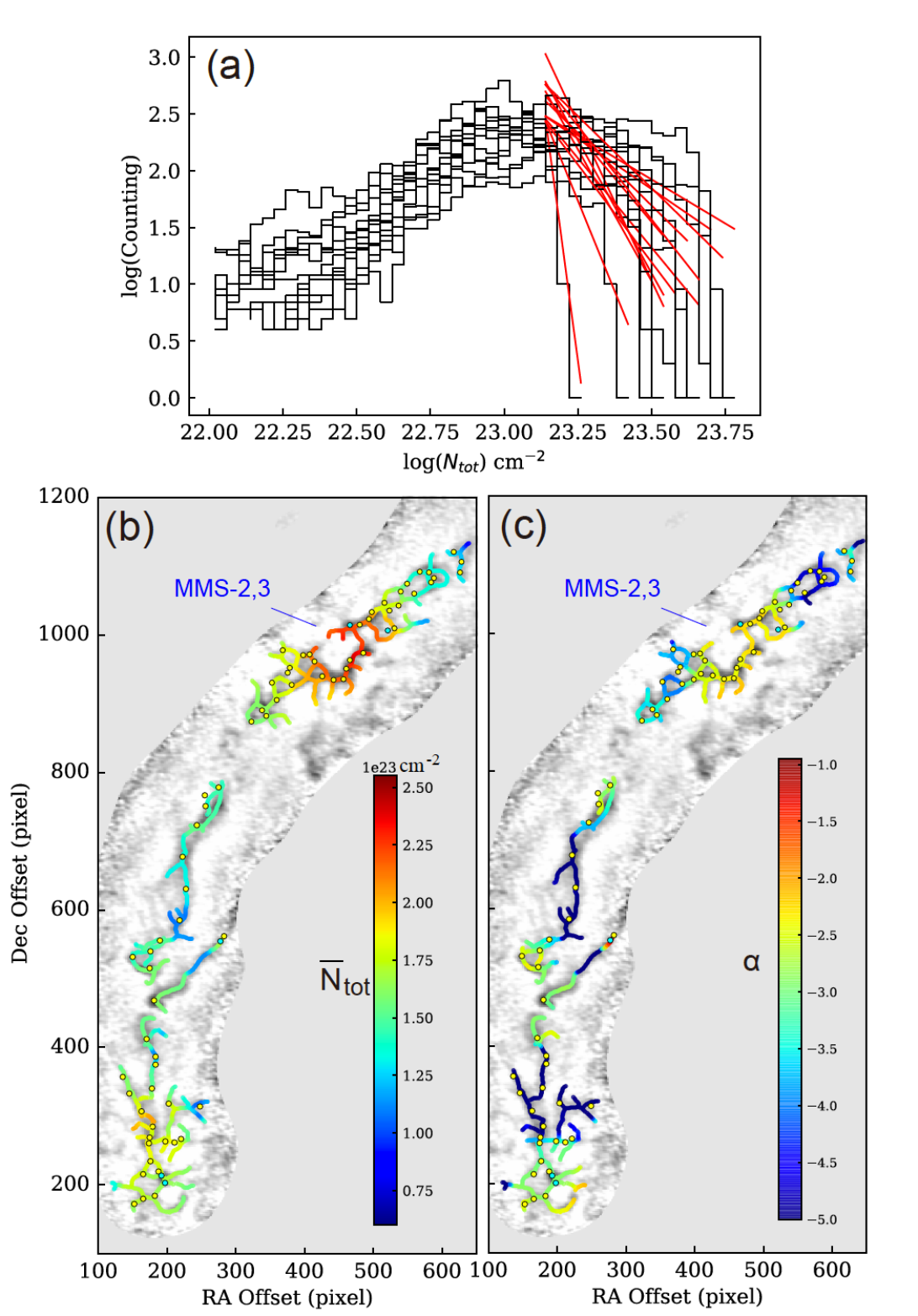}
	\caption{{\bf (a)} A number of N-PDF profiles overlaid to show the N-PDF variation. The red lines represent the power-law fitting to each N-PDF in the range of $\log N_{\rm tot}>23.2$ cm$^{-2}$. At each point, the N-PDF is measured in a circular area with $r=14$ arcsec. {\bf (b)} The average $N_{\rm tot}$ distribution over the filament paths in OMC-3. {\bf (c)} The distribution of N-PDF power-law tail index $\alpha$ over the filament paths in OMC-3. The $\overline{N}_{\rm tot}$ and $\alpha$ distributions are measured over the filament paths with a spatial interval of $1.3$ arcsec (nearly half beam size) in order to keep a uniform Nyquist spatial sampling.}                                         
	\label{fig:npdf_map}
\end{figure*}

\begin{figure*}
	\centering
	\includegraphics[width=0.9\textwidth]{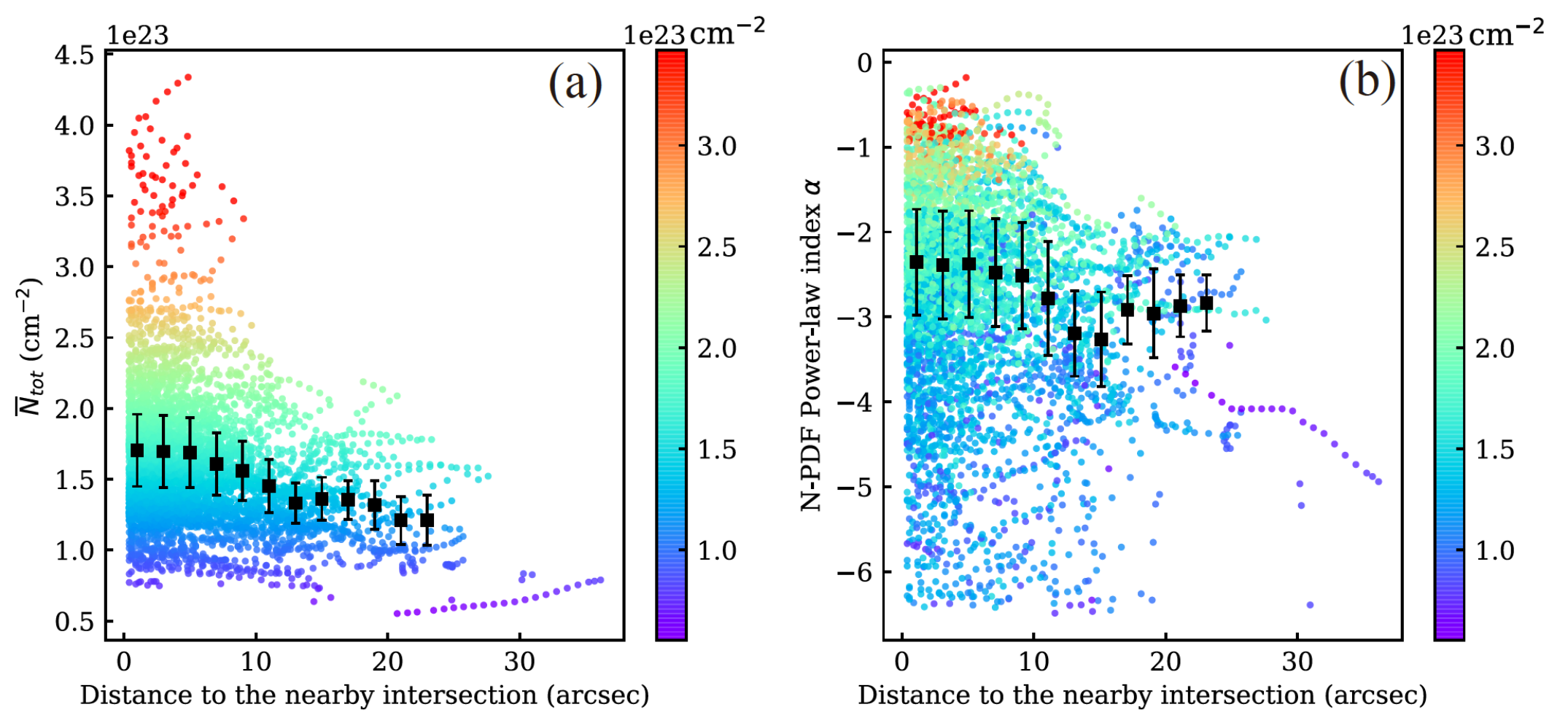}
	\caption{The distribution of the parameters with the distance from each sampled position to its nearest intersection, $d_i$, for the positions over the filament paths in OMC-2,3 (see Fig.\,\ref{fig:npdf_map}b and \ref{fig:npdf_map}c for OMC-3). The squares with the error bars represent the average value and standard deviation within each interval, which has a width of $\Delta d_i=2$ arcsec. {\bf (a)} The distribution of $\overline{N}_{\rm tot}$ with $d_i$. {\bf (b)} The distribution of $\alpha$ with $d_i$. In both (a) and (b), the color scale represents $\overline{N}_{\rm tot}$. }                                         
	\label{fig:npdf_stat}
\end{figure*}

\begin{figure*}
	\centering
	\includegraphics[width=0.9\textwidth]{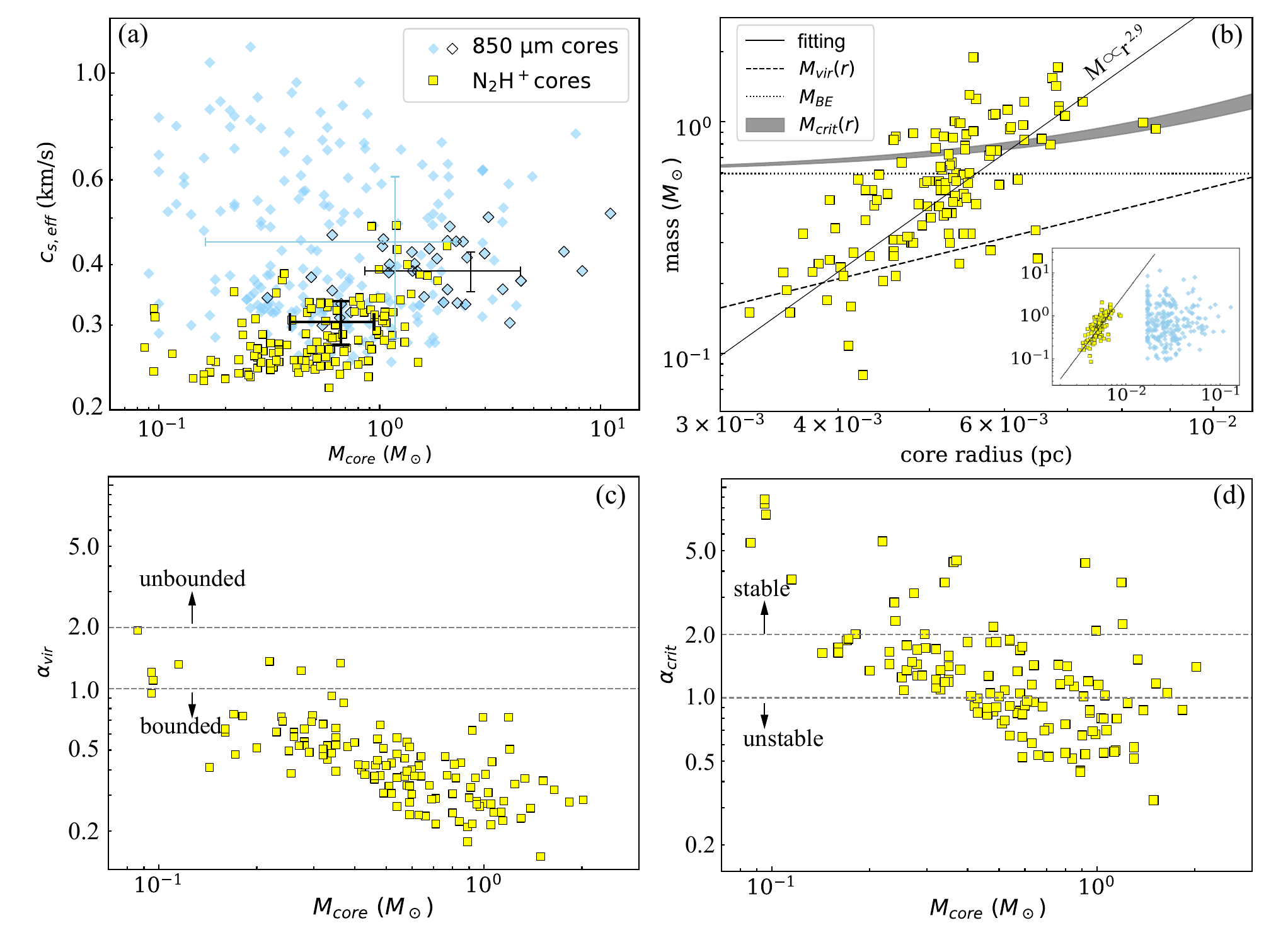}
	\caption{ The distributions of the core parameters. {\bf (a)} The diagram of effective sound speed $c_{\rm s,eff}$ and the core mass $M_{\rm core}$ for the $\nthp$ cores. The yellow squares represent the $\nthp$ cores in this work, while the diamonds represent the 850 $\micron$ dusty cores \citep{kirk17} whose $c_{\rm s,eff}$ values are calculated from $\sigma_{\rm obs}$ and $T_{\rm kin}$ values (Table 1) using Equation \ref{equ:cs_eff2}. The 850 $\micron$ cores are sampled over a spatial scale of 12pc, with Orion BN/KL and Orion A South regions also included. In \citet{kirk17} a spatially coherent GBT $\nht$ observation was also performed to measure $\sigma_{\rm obs}$ and $T_{\rm kin}$ of the cores. Among the 850 $\micron$ cores, the black-edged diamonds represent those overlapped with the $\nthp$ emission, thus the possible parental cores of the $\nthp$ filaments and cores. The solid error bars represent the average values and standard deviation of the three samples, $\nthp$ cores (thick black lines), the 850 $\micron$ cores associated with the $\nthp$ emission (thin black lines), and the entire sample of the 850 $\micron$ cores (blue lines). {\bf (b)} The distribution of the core mass and radius. The core radius represents the average value of the major and minor axes. The solid line represents the power-law fitting to the data points. The virial mass and BE mass are plotted in dashed and dotted lines, respectively. The gray strip represents the $M_{\rm crit}(r)$ range due to the magnetic field variation between 0.64 to 0.85 mG \citep{houde04,matthews05}. The insert panel shows the data 850 $\micron$ dusty cores in blue diamonds. In calculation of $M_{\rm vir}(r)$ and $M_{\rm crit}(r)$, an average value of $c_{\rm s,eff}=0.3$ \kms\ is assumed. {\bf (c)} The diagram of $M_{\rm core}$ and $M_{\rm vir}$. The dashed lines represent the levels of $\alpha_{\rm vir}=1$ and 2. {\bf (d)} The distribution of $M_{\rm core}$ and $M_{\rm crit}$. The dashed lines also represent $\alpha_{\rm crit}=1$ and 2, respectively. In (c) and (d), $M_{\rm vir}$ and $M_{\rm crit}$ are calculated from the $c_{\rm s,eff}$ values of the individual cores instead of using the average values. }                                          
	\label{fig:r_mvir}
\end{figure*}

\label{lastpage}
\end{document}